\documentclass[12pt,epsf,epsfig,aps,floatfix,nofootinbib,superscriptaddress]{revtex4-2}
\usepackage{graphics}
\usepackage{graphicx}
\usepackage[english]{babel}
\usepackage{amssymb}
\usepackage{amsmath}
\usepackage{xcolor}
\usepackage{hyperref}

\newcommand{\nc}{\newcommand}
\newcommand{\be}{\begin{equation}}
\newcommand{\ee}{\end{equation}}
\newcommand{\beq}{\begin{eqnarray}}
\newcommand{\eeq}{\end{eqnarray}}
\nc{\barray}{\begin{eqnarray}}
\nc{\earray}{\end{eqnarray}}
\nc{\barrayn}{\begin{eqnarray*}}
\nc{\earrayn}{\end{eqnarray*}}
\nc{\bcenter}{\begin{center}}
\nc{\ecenter}{\end{center}}
\nc{\ket}[1]{| #1 \rangle}
\nc{\bra}[1]{\langle #1 |}
\nc{\0}{\ket{0}}
\nc{\mc}{\mathcal}
\nc{\er}[1]{(\ref{eq:#1})}
\nc{\onehalf}{\frac{1}{2}}
\nc{\partialbar}{\bar{\partial}}
\nc{\psit}{\widetilde{\psi}}
\nc{\Tr}{\mbox{Tr}}
\nc{\eV}{\;\mathrm{eV}}
\nc{\MeV}{\;\mathrm{MeV}}
\nc{\GeV}{\;\mathrm{GeV}}
\nc{\TeV}{\;\mathrm{TeV}}

\def\chii0{\chi_i^0}
\def\chij0{\chi_j^0}

\newcommand{\gsim}{\lower.7ex\hbox{$\;\stackrel{\textstyle>}{\sim}\;$}}
\newcommand{\lsim}{\lower.7ex\hbox{$\;\stackrel{\textstyle<}{\sim}\;$}}

\usepackage{stackengine}
\stackMath

\begin{document}

\title{Dark Matter Candidates of a Very Low Mass}
\author{Kathryn M. Zurek}

\begin{abstract}

We review dark matter (DM) candidates of a very low mass appearing in the window below the traditional weakly interacting massive particle ($m_\chi \lesssim 10$~GeV) and extending down to $m_\chi \gtrsim 1$~meV, somewhat below the mass limit at which DM becomes wavelike.  Such candidates are motivated by hidden sectors such as hidden valleys, which feature hidden forces and rich dynamics, but have evaded traditional accelerator searches for New Physics because of their relatively weak coupling to the Standard Model (SM).  Such sectors can still be detected through dedicated low-energy colliders, which, through their intense beams, can have sensitivity to smaller couplings, or through astrophysical observations of the evolution of DM halos and stellar structures which, through the Universe's epochs, can be sensitive to small DM interactions.  We also consider mechanisms whereby the DM abundance is fixed through the interaction with the SM, which directly motivates the search for light DM in terrestrial experiments.  The bulk of this review is dedicated to the new ideas that have been proposed for direct detection of such DM candidates of a low mass through nuclear recoils, electronic excitations, or collective modes such as phonons and magnons.  The rich tapestry of materials and modes in the condensed matter landscape is reviewed along with specific prospects for detection.

\end{abstract}

\maketitle
\newpage

\tableofcontents

\section{Hidden Sector Dark Matter: An Introduction}

In the search for DM candidates, a few considerations enter---most notably that any DM candidate must satisfy observational evidence.  This begins with the observed DM density, which is fixed most precisely at the epoch of the cosmic microwave background (CMB, when the Universe was approximately 380,000 years old, at a redshift of $z \sim 10^3-10^4$).  From that epoch until today, it is known---on the largest scales of the Universe, from the formation of galaxies and clusters of galaxies as well as galactic rotation curves---that DM density ($\rho$) must dilute with the expanding volume of the Universe, $\rho = \rho^0(1+z)^3$ (with $\rho^0$ the density today), and must have very weak interactions with the baryons (so as not to disturb CMB baryon acoustic oscillations) and also with itself so that the structure of galaxies remains approximately oblate (neither a perfect sphere nor a disk).  That is, {\em on average}, the DM must have a cold equation of state ($w = 0$) at least since the time the Universe was approximately 380,000 years old and must have weak enough interactions with itself and with ordinary matter that the gravitational force dominates the formation of bound structures, and not other forces like baryonic pressure.  Beyond this broad set of facts, little is known about the DM.  For a classic review of the evidence for DM, we refer the reader to, for example, Reference~\cite{Bertone:2004pz}.

Around the time that these facts about the Universe were becoming broadly accepted (in the 1980s), particle physics was itself struggling with some puzzles that mostly concerned questions of naturalness.  It turned out that solving these problems of naturalness in the Standard Model (SM) of particle physics naturally produced two DM candidates.  The first of these candidates is the weakly interacting massive particle (WIMP), which appears as part of the solution to the question of why the Higgs boson mass (and hence weak forces) is so light; by extension, the WIMP mass is typically at the weak scale, and WIMPs have weak interactions, making them very susceptible to detection with the barrage of experiments that probe the weak scale.  The second of these candidates is the axion, a possible solution to the question of why there is so little CP violation in the strong interactions.  The axion has interactions that are much weaker than weak (corresponding to a mass scale of $10^9$~GeV up to the Planck scale), but its much lower mass, and hence much higher number density, gives a boost to the detection probability, typically in electromagnetic cavities that exploit the coherent enhancement of a wavelike state such as axion DM. Both WIMPs and axions could be produced in the early Universe (thermally and non-thermally, respectively) in an abundance consistent with observations.  Multiple experimental efforts are underway to search for these DM candidates.  Because they are part of the solution to the SM's problems, they have interactions with the SM that allow one to predict detection in a sufficiently sensitive device with ordinary particles.  Well-defined predictions are appealing to the intrepid DM hunter. We refer the reader to recent reviews (\cite{Cooley:2022ufh,Adams:2022pbo}) that describe these ongoing efforts.

However, the narrow focus on these two candidates sidesteps the fact that, from an observational point of view, a huge mass range of DM particles and almost as large a range of interactions are observationally possible.    On the low mass end, DM can be as light as is consistent with the formation of structure on dwarf-galaxy scales, which implies that the de Broglie wavelength of the DM must be shorter than a typical dwarf galaxy size, implying $m_\chi \gtrsim 10^{-22}$~eV.  On the upper mass end, DM is not observed to be clumpy (or grainy) in measurements of the Lyman-$\alpha$ forest; this implies that DM should have a mass $m_\chi \lesssim 10^3~M_\odot$ (for reference, $1~M_\odot \sim 10^{57}$~GeV) to be sufficiently smooth (see Reference~\cite{Afshordi:2003zb}, in which shot noise fluctuations from primordial black hole DM were considered).  Any mass in-between is consistent with observations.

Once one releases the requirement that the DM solve one of the SM's problems, a hidden world of possibilities opens, subject only to the requirement that the DM 
\begin{itemize}
\item have the observed abundance,
\item  dilute as a cold non-relativistic state after the CMB epoch,
\item have its interactions on large scales be dominated by gravity, and
\item satisfy the requirement that DM-baryon interactions not damp baryons after the CMB epoch and not affect the formation of nuclei in the early Universe.
\end{itemize}
In fact, there is not even a requirement that the hidden world have only one dark state, that it have self-interactions much weaker than baryonic, or that {\em all} of the dark sector be nonrelativistic.  

\begin{figure}
\includegraphics[width=0.75\linewidth]{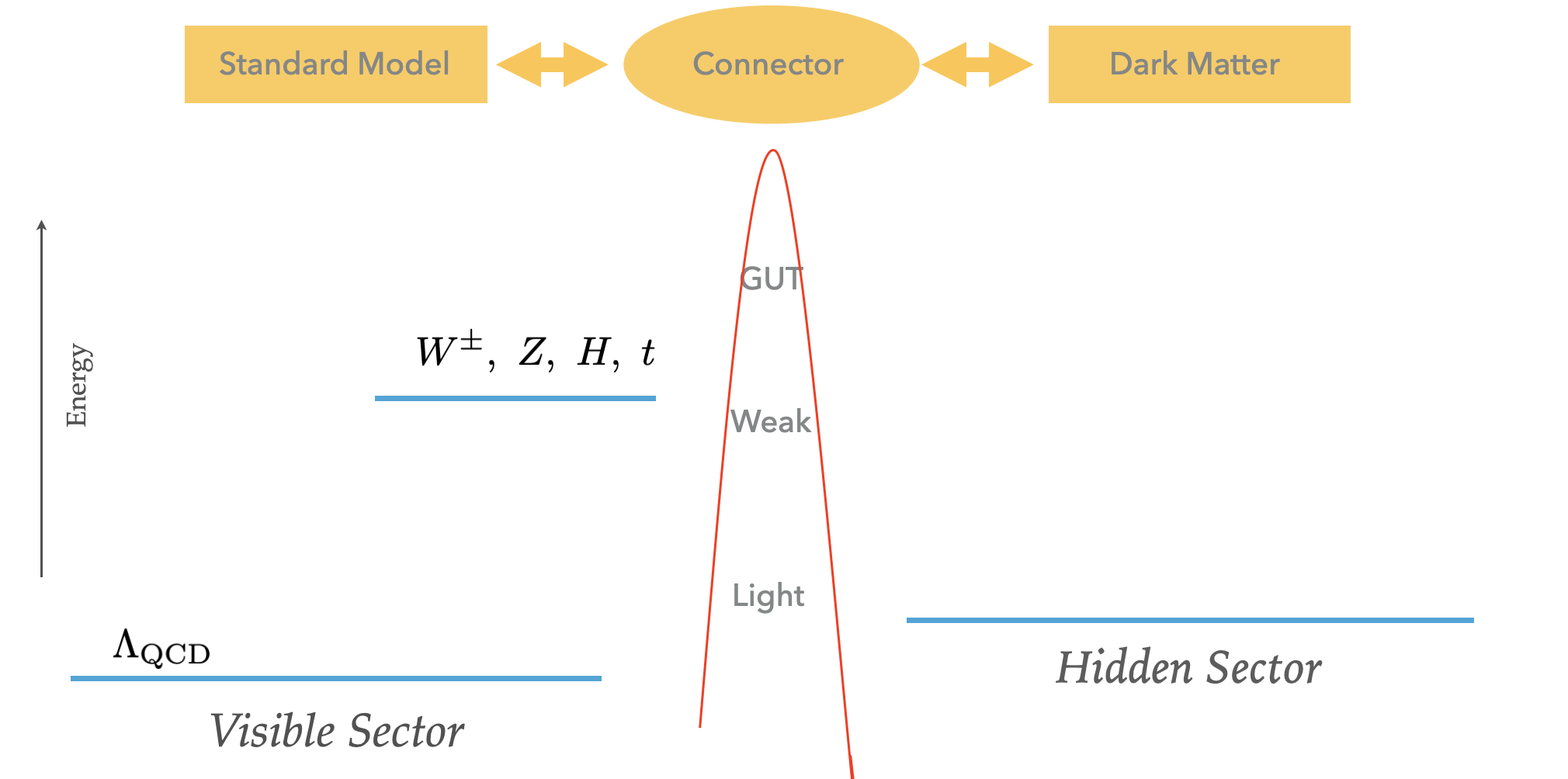}
\caption{Schematic of hidden-sector dark matter.  The barrier in the center of the figure represents the interaction (from a GUT, weak scale, or light mediators) between the two sectors; a higher barrier represents a weaker interaction.  The mediator connects the visible and hidden sectors, and the different heights of the floors in the two sectors reflect the mass gap.  In the visible sector we have the Standard Model, with a mass gap for the baryons of around a GeV, while in the dark matter sector, the scale of the mass gap and the structure of the states there are unknown. Abbreviation: GUT, grand unified theory.
\label{fig:1}
}
\end{figure}

The structure of a hidden sector (also known as a hidden valley or dark sector)  is shown schematically in Figure~\ref{fig:1}.  The barrier between the SM and the hidden sector represents the interaction between the two sectors; a higher barrier represents a weaker interaction.  The highest barrier is the weakest force of all, gravity.  To detect DM through individual particle interactions with the SM, there must be other, much lower barriers; these are represented as lower peaks.  The mass scale of the hidden sector, represented by the height of the ``floor,'' is unknown.

One is tempted to become overwhelmed by such a huge mass range of DM, the types of dark sectors, and their interactions.  Thus, it helps to break things down according to the physics governing the DM masses and interactions.  
In this article, we focus on a mass range of particle DM that is motivated, by its relic abundance, to have large enough interactions with the SM to be detectable through particle interactions.  When the DM is heavier than approximately 10 TeV, setting its relic abundance through interactions with the SM is challenging (due to constraints from perturbative unitarity, as explicitly shown below).  
For heavier-mass DM, one can use gravitational means ({\em e.g.} pulsar timing~\cite{Ramani:2020hdo} or the unique motions of stars~\cite{VanTilburg:2018ykj}) or an additional DM-SM fifth force to search for such candidates~\cite{Gresham:2022biw}, often through astrophysical means.  On the other end of the mass scale, when the DM is lighter than $\sim 1$~eV, it behaves more like a wave than an individual particle.  Detection techniques in this ultralight mass regime focus on using coherence, often in electromagnetic cavities or with other AMO techniques~\cite{Battaglieri:2017aum}.  While this is a vibrant area of research, it is not the focus of this article.

We focus on hidden-sector DM candidates in the low mass range, whose relic abundance is still naturally set by its interactions with the SM, where there is motivation to search for such a state through detection of individual particles in terrestrial experiments.  This implies a DM mass range between approximately a few keV (below which DM that is produced by its particle interactions with the SM will be too warm to cluster appropriately; see {\it e.g.} Reference~\cite{DES:2020fxi}) and approximately 10 GeV, just below the weak scale.  Because these hidden-sector states have masses below the weak-scale interactions, making for a low mass floor in the schematic of Figure~\ref{fig:1}, we refer to these DM models as hidden-valley DM (HVDM) or hidden-sector DM (HSDM).  

The outline of the rest of this review is as follows.  In the next section, we review how DM candidates of a very low mass can naturally have their relic abundance set through interactions with the SM.  We characterize the various kinds of mechanisms that are often used in the literature to set the relic abundance.  We then turn to examining astrophysical and cosmological probes for HVDM, a discussion that by itself sheds a light on where DM might be detected in terrestrial experiments.  Then we turn to the main subject of the article, terrestrial probes for HVDM, and focus mostly on direct detection experiments.  Novel probes must be invented to search for HVDM because the traditional nuclear recoil probes of WIMP DM are ineffective.  More specifically, this implies using the wealth of collective modes (such as phonons and magnons) that are available in condensed matter systems.  We also briefly review other kinds of terrestrial probes for HVDM, such as accelerator-based experiments.

\section{Models of Light Dark Matter: a Brief Overview}

The broad range of hidden-sector models necessitates a tight selection of theory for this review.  We focus here on direct detection prospects in terrestrial experiments and, therefore, on interactions with the SM that could be detected.  These interactions are the most motivated when they also set the relic abundance, as discussed in the next section.  In the dark sector itself, the structure is relatively unconstrained, especially if the mass gap in the hidden sector is $\gtrsim 10$~MeV, where constraints from big bang nucleonsynthesis (BBN) from a thermalized hidden sector can be most easily satisfied.  For example, the hidden sector could be~\cite{Strassler:2006im,Strassler:2006ri}
\begin{itemize}
\item a QCD-like theory with $F$-flavors and $N$-colors, with only light or heavy quarks, or adjoint quarks;
\item a QED-like theory with only massive force mediators;
\item a pure-glue theory;
\item a remnant from supersymmetry breaking;
\item a partially Higgsed SU($N$) theory;
\item a Seiberg duality cascade;
\item unparticles; or
\item a sector arising from a Randall--Sundrum or Klebanov-Strassler throat in extra dimensions.
\end{itemize}
Hidden sectors with such structures will naturally give rise to DM self-interactions.  Strongly coupled hidden sectors have weakly coupled duals, so it is also possible to write effective Lagrangians in terms of weakly coupled scalars, fermions, and vector mediators~\cite{Boehm:2003hm}.  To take a particularly simple example, the dark sector could be a QCD-like theory with two colors, two light flavors, and no heavy flavors~\cite{Strassler:2006im}.  Such a hidden sector has gapped degrees of freedom and the hidden pions 
$\pi^\pm_v$ and $\pi_v^0$. 
Note that the plus-minus symbol here does not denote electric charge but rather charge under a new global symmetry in the hidden sector that stabilizes the pions carrying the global symmetry $\pi^\pm_v$.  The relic abundance is fixed by the rate of the annihilation process
\beq
\pi^+_v \pi^-_v \rightarrow \pi_v^0 \pi_v^0,
\eeq
with $\pi_v^0 \rightarrow f \bar f$ decaying, for example, to SM fermions to which the dark-sector states couple via the connector state.  This structure also appears in secluded DM models~\cite{Pospelov:2007mp,Pospelov:2008jd}.  

Note that it is quite natural in such theories for the DM to have an asymmetric relic abundance, as proposed in Reference~\cite{Kaplan:2009ag}.     For example, a large enough asymmetry (or chemical potential) between $\pi_v^+$ and its anti-particle in the example above changes the cosmology and the computation of the relic abundance, as discussed in more detail below.  The reader also may consult Reference~\cite{Zurek:2013wia} for a discussion of HVDM/HSDM in the context of asymmetric DM.  This general framework of asymmetric DM triggered a slew of model building, from atomic~\cite{Kaplan:2009de}, quirky~\cite{Kribs:2009fy} and glueball~\cite{Boddy:2014yra} DM to the WIMPless~\cite{Feng:2008ya} and strongly interacting massive particle (SIMP)~\cite{Hochberg:2014kqa} ``miracles".

Hidden sectors also naturally allow for ``portals'' between the two sectors, through which the DM interacts with the SM, giving rise to signatures in terrestrial experiments.  Since the DM is electrically neutral, there are two natural portals involving dimension-4 operators:
\begin{itemize}
\item A vector portal that includes a dark photon $A'_\mu$ kinetically mixed with the photon~\cite{Holdom:1985ag} (see Reference~\cite{Hooper:2008im} for an application to DM):
\beq
{\cal L}_{A'} \supset -\frac{1}{2} m_{A'}^2  - \frac{1}{4} {F'}^{\mu \nu} F'_{ \mu \nu} - \frac{\epsilon}{2} F_{\mu \nu} {F'}^{\mu \nu} - y_\chi A'_\mu \bar \chi \gamma^\mu \chi.
\label{eq:VectorPortal}
\eeq
The vector mediator $A'_\mu$ can also, for example, be a gauge boson from a $B-L$ symmetry, $U_{B-L}$~\cite{Knapen:2017xzo}.
\item A scalar portal that includes a scalar mixing with the Higgs boson~\cite{Strassler:2006ri,Patt:2006fw}:
\beq
{\cal L}_{H \phi} \supset -\frac{1}{2} m_\phi^2 |\phi|^2- \kappa |H|^2 |\phi|^2.
\label{eq:ScalarPortal}
\eeq
One can also have a hadrophilic or leptophilic scalar (as discussed in detail in Reference~\cite{Knapen:2017xzo}) or an axion or conformal portal.  
\item A neutrino portal is also often considered, via the SM-neutral combination of fields $L H$, where $L$ is an SM lepton doublet:
\beq
{\cal L}_N \supset L H N + g N \chi \phi + m_N N N.
\eeq
We refer the reader to Reference~\cite{Hooper:2004dc} and \cite{Kaplan:2009ag} for early examples and to Reference~\cite{Alekhin:2015byh} for its implementation.  This portal does not typically appear in simple models for direct detection of light DM, though it is important for models of asymmetric DM, which we discuss below.
\end{itemize}

Another important feature in the hidden sector, besides its self-interactions and the portal to the visible sector, is the mass gap fixing $m_\chi$.  While by no means necessary, the theory becomes more predictive if this gap is fixed by a relation with the visible sector.  This can happen, for example, if the confinement scale in the hidden sector is triggered by visible sector confinement~\cite{Murgui:2021eqf} or if supersymmetry breaking  in the visible sector triggers supersymmetry breaking in the hidden sector and fixes the mass scales in that sector~\cite{Cohen:2010kn}.  While we do not go into the details of these models in this review, the mechanism that generates the mass gap in the hidden sector can also give rise to predictive interaction rates in terrestrial experiments; we explore some of these cases in the next two sections.

\section{Mechanisms for Light Dark Matter Relic Abundance}

The direct search for DM candidates whose masses are below the electroweak scale depends on the DM having couplings to the SM other than the gravitational one.  The strength of those couplings, and in particular whether they are large enough to give rise to detectable signals in terrestrial experiments, is best motivated if the DM relic abundance is fixed through its interactions with the SM.  We review here mechanisms for setting the relic abundance; our goal is to show how such DM candidates could be observed through a scattering or absorption process in a direct detection experiment.

\subsection{Simplified Models in Direct Detection}

\begin{figure}
\includegraphics[width=0.55\linewidth]{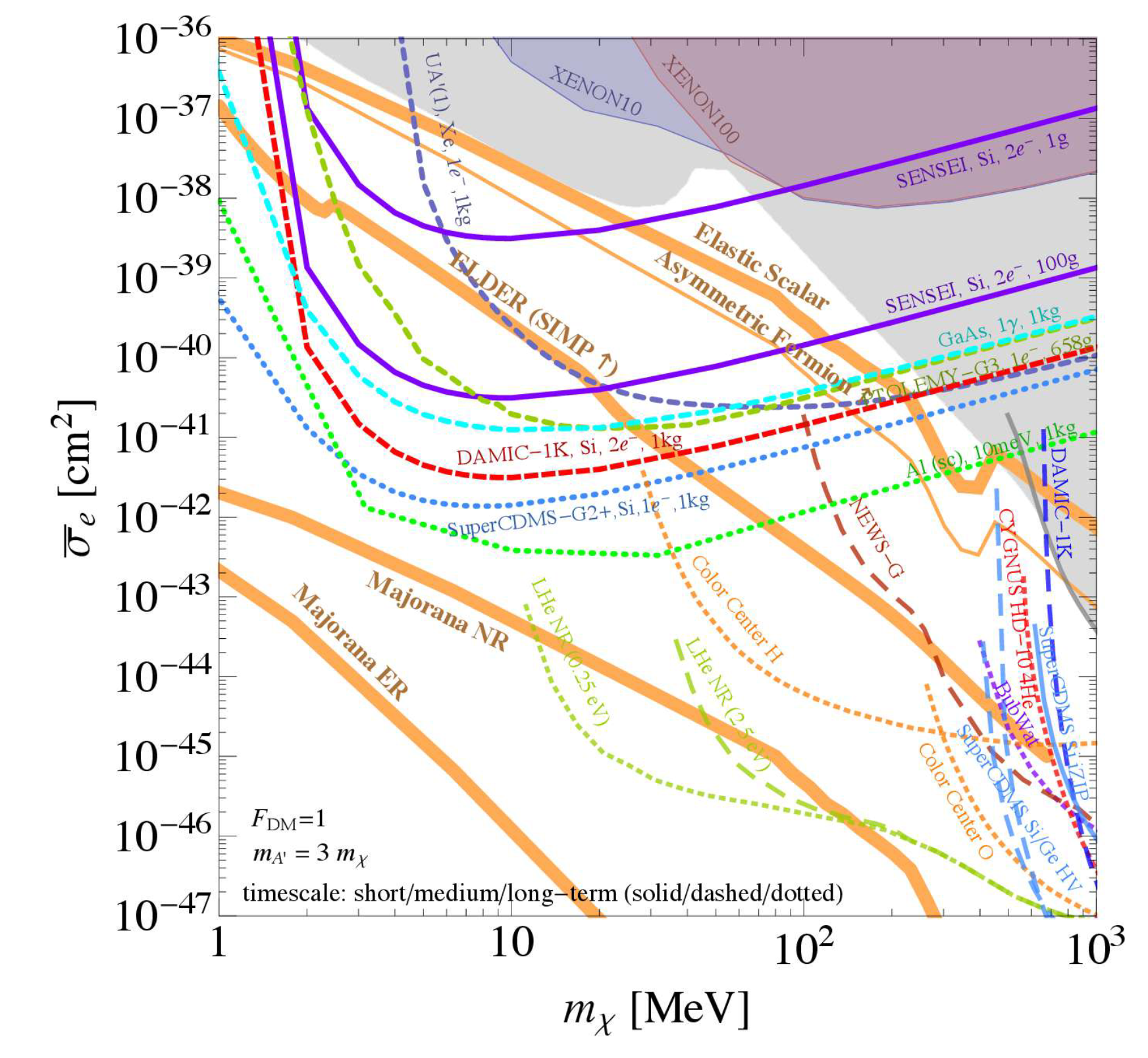}
\caption{Direct detection cross section ($\bar \sigma_e$) reach for a massive vector mediator coupled to electrons as a function of DM mass ($m_\chi$).  Orange bands reflect five sets of model assumptions and correspond to the combination of couplings that fix the relic abundance through {\em (i)} $P$-wave freeze-out of scalar DM (elastic scalar; see discussion around Equation~\ref{eq:yfoswave}), {\em (ii)} asymmetric fermion DM (see discussion around Equation~\ref{eq:CMBbound}), or {\em (iii)} ELDERs/SIMPs (see Equation~\ref{eq:yELDER}).  These cases motivate searches by direct detection.  If the DM is a Majorana fermion (Majorana NR or Majorana ER), its direct detection cross sections are suppressed and accelerator searches are better equipped for detection. Various ideas have been proposed to directly detect light DM, such as through electronic excitation in a silicon (SENSEI, DAMIC, SuperCDMS) or GaAs target, superconductors (Al SC), color centers, superfluid helium, or polar materials. Solid lines indicate a short-term timescale, dashed lines indicate a medium-term timescale, and dotted lines indicate a long-term timescale.  The shaded regions indicate existing constraints. For more details regarding the experimental proposals, see Reference~\cite{Battaglieri:2017aum}. Abbreviations: Al SC, superconducting aluminum; DM, dark matter; ELDER, elastically decoupling relic; ER, electron recoil; HV, high voltage; NR, nuclear recoil; SIMP, strongly interacting massive partcile. Figure adapted from Reference~\cite{Battaglieri:2017aum}.
\label{fig:DDspacefreezeout}
}
\end{figure}

Equations~\ref{eq:VectorPortal},~\ref{eq:ScalarPortal} comprise simplified models of portals mediating interactions with the SM, where $m_\phi$ and $m_{A'}$ can be denoted more generally as mediators of mass $m_M$.
It is often useful to employ simplified models as a general framework for understanding how relic abundance considerations interplay with astrophysical, cosmological, and direct detection constraints.  A typical interaction cross section of DM with the coupling $\alpha_\chi = \frac{g_\chi^2}{4 \pi}$ via a mediator, on a target $T$ (electron or nucleon) having a coupling $\alpha_T = g_T^2/4 \pi$,  takes the form 
\beq
\bar \sigma_{T} \equiv \frac{16 \pi \alpha_T \alpha_\chi}{(m_M^2 + {\bf q}_0^2)^2} \mu_{T \chi}^2
\label{eq:DDreference}
\eeq
in a direct detection experiment, where $\mu_{T \chi}$ is the target-DM reduced mass and ${\bf q}_0$ is a typical momentum transfer in the experiment, used here to define the reference cross section.  For electron targets, the reference momentum is taken to be $q_0  = \alpha m_e$. The same product of couplings, $\alpha_T \alpha_\chi$, typically sets the relic abundance of DM in models where DM is produced through its interaction with the SM.  In the case of scattering via a heavy vector mediator on an electron target, Equation~\ref{eq:DDreference} with $T = e$ can be written as
\beq
\bar \sigma_e \equiv \frac{16 \pi \alpha_e m_e^2}{m_\chi^4} y_{\rm RA},
\label{eq:eDD}
\eeq
where we have taken $m_\chi \gg m_e$ and separated a relic abundance parameter $y_{RA}$:
\beq
y_{\rm RA} \equiv \alpha_\chi \left( \frac{m_\chi}{m_{A'}} \right)^4.
\label{eq:yRA}
\eeq

\begin{figure}
\includegraphics[width=0.55\linewidth]{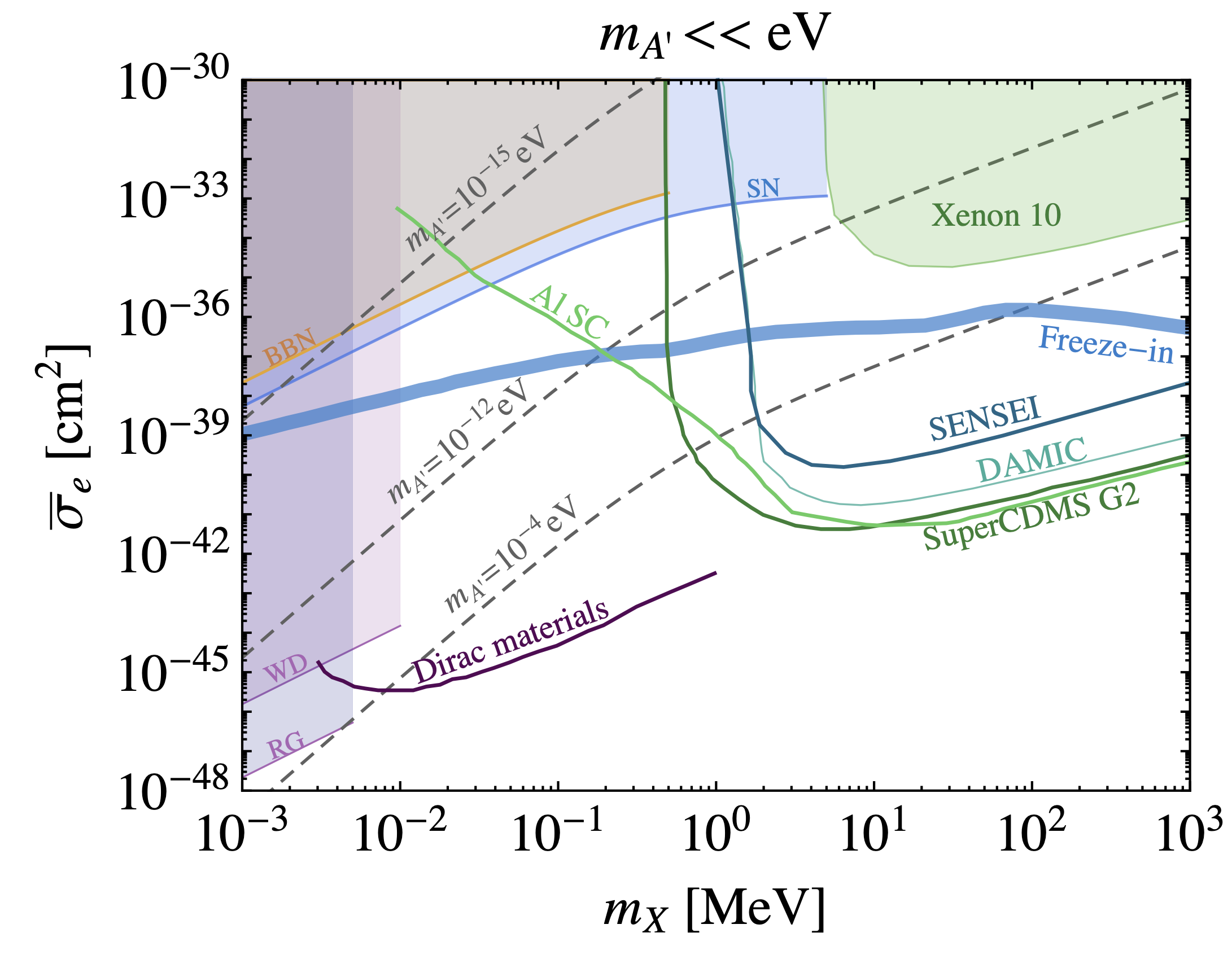}
\caption{Direct detection cross section ($\bar \sigma_e$) reach for a massless dark photon mediator as a function of DM mass ($m_\chi$).  The blue band corresponds to the combination of couplings (Equation~\ref{eq:freezeingchige}) where the observed relic abundance is fixed by freeze-in.  Various ideas have been proposed to directly detect light DM, such as through a silicon or germanium target (SENSEI, DAMIC, SuperCDMS), superconductors (Al SC), Dirac materials, or polar materials.  The shaded regions indicate constrained parameter space, such as bounds provided by stellar cooling (including RGs, WDs, and SNe) as well as by BBN. Abbreviations: Al SC, superconducting aluminum; BBN, big bang nucleosynthesis; DM, dark matter; RG, red giant; SN, supernova; WD, white dwarf. Figure adapted from Reference~\cite{Knapen:2017xzo}.
\label{fig:DDspacefreezein}
}
\end{figure}

Summary plots showing the proposed reach in $\bar \sigma_e$  of several direct detection experiments as a function of light DM mass are shown in Figure~\ref{fig:DDspacefreezeout} for a massive mediator and in Figure~\ref{fig:DDspacefreezein} for a massless mediator.  In the rest of this review, one goal is to illuminate how the various reach curves and constraints are computed.  The orange bands in Figure~\ref{fig:DDspacefreezeout} and the blue band in Figure~\ref{fig:DDspacefreezein} correspond to the regions of model space where the DM abundance in the Universe is fixed by the same couplings that give rise to the scattering in direct detection experiments.  We now turn to discussing concretely how the scattering cross section in direction detection experiments can be related to the relic abundance in three common cases that set benchmarks in terrestrial searches for light DM.  

\subsection{Thermal Freeze-Out of Symmetric Dark Matter}
\label{subsec:freezeout}

Thermal freeze-out occurs when the temperature drops below the DM mass and the equilibrium process, $\chi \bar \chi \leftrightarrow f \bar f$, becomes simply an annihilation process,
\beq
\chi \bar \chi \rightarrow f \bar f,
\eeq
since the DM is too heavy to produce through the inverse process $f \bar f \rightarrow \chi \bar \chi$ when $T \ll m_\chi$.
Freeze-out fixes the relic abundance to a value obtained by solving a Boltzmann equation,
\beq
\frac{dY_\pm}{dx} = - \frac{x \langle \sigma v\rangle s}{H(m_\chi)}(Y_+ Y_- - (Y^\mathrm{eq})^2),
\eeq
where we are allowing DM to have separate number densities for particle and antiparticle to permit a DM asymmetry, and $Y_\pm = n_\pm/s$, where $s$ is the entropy density.  Here $x = m_\chi/T$ is a measure of temperature (or time), the Hubble parameter $H(m_\chi)$ is evaluated at $T = m_\chi$ in a radiation-dominated universe, and $Y^\mathrm{eq} \simeq a x^{3/2} e^{-x}$ is the equilibrium number density, with $a$ dependent on the number of thermalized degrees of freedom. The annihilation cross section is parameterized as S-wave ($n = 0$) or P-wave ($n=1$):
$
\langle \sigma v\rangle \equiv \sigma_0 (T/m)^n = \sigma_0 x^{-n},
$
where $\frac{3}{2} T = \frac{1}{2} m \langle v^2\rangle$.  Note that for P-wave interactions, since the DM velocity drops as the Universe cools, the interaction rate is much smaller in the late Universe than in the early. This can be important for understanding constraints on the annihilation cross section from later epochs---for example, from the CMB at a temperature of approximately 1 eV.  In the case where there is no DM particle--antiparticle asymmetry, freeze-out occurs when $n_\chi \langle \sigma_A |v|\rangle \simeq H(x_f)$, and one can do a back-of-the-envelope calculation of the needed annihilation cross section to fix the observed relic abundance $\rho_\chi^0$:
\begin{eqnarray}
\langle \sigma v\rangle & \sim & \frac{T_f^2}{M_\mathrm{Pl} }\frac{m_\chi}{\rho_\chi(T_f)} \\ \nonumber
& = & \frac{T_f^2}{M_\mathrm{Pl} }\frac{m_\chi}{\rho_\chi^0}\frac{T_0^3}{T_f^3} \\  \nonumber
& = & \frac{T_0^3}{M_\mathrm{Pl} }\frac{x_f}{\rho_\chi^0} \\ \nonumber
& \simeq & 3 \times 10^{-26} \mbox{ cm}^3 \mbox{ s}^{-1} \left(\frac{x_f}{20}\right), \nonumber
\end{eqnarray}
where $x_f\simeq 20$ is the freeze-out temperature obtained from solving the Boltzmann equation for S-wave annihilation and DM in the WIMP window $m_\chi \simeq 1$~TeV.  It has only logarithmic sensitivity to the DM mass.  An S-wave annihilation process to a light vector particle scales parametrically as
\beq
\langle \sigma v \rangle \simeq \frac{\pi \alpha_\chi^2}{ m_\chi^2} \sqrt{1-\left(\frac{m_M}{m_\chi}\right)^2}  \simeq 3 \times 10^{-26} \frac{\mbox{cm}^3}{\mbox{s}} \left(\frac{g_\chi}{0.4}\right)^4 \left(\frac{2 \mbox{ TeV}}{m_\chi}\right)^2,
\label{eq:FOxsec}
\eeq
 where we have taken the vector force mass to be much smaller than the DM mass, $m_M \ll m_\chi$.  The fact that $g_\chi \sim 1$ when $m_\chi \sim 1 \mbox{ TeV}$ is parametrically the reason that weak-scale DM is said to be motivated by thermal freeze-out, a statement known colloquially as the WIMP miracle.
 
 One can see immediately that the DM cannot be pushed much heavier than $\sim 10 \mbox{ TeV}$ without running into a regime where couplings become non-perturbative and the theory is inconsistent.  However, as DM becomes lighter, one simply needs to scale down the coupling product $\alpha_\chi$ linearly with $m_\chi$ to satisfy the same relic abundance considerations.  Thus, HSDM of a low mass is equally well motivated by thermal freeze-out considerations.  
 
 More generally, for an $s$-channel annihilation process to electrons through a mediator, vector or scalar, the annihilation cross sections at freeze-out are
 \beq
 \langle \sigma v \rangle_V \simeq \frac{ 16 \pi \alpha_\chi \alpha_e m_\chi^2}{(4 m_\chi^2 - m_{A'}^2)^2}~~~ \mbox{          and          } ~~~\langle \sigma v \rangle_S \simeq \frac{ 2 \pi \alpha_\chi \alpha_e m_\chi^2}{(4 m_\chi^2 - m_{S}^2)^2}\frac{1}{x_f}.
 \label{eq:annxsec}
 \eeq
As DM drops below approximately 10 GeV down to $2 m_e$, its relic number density remains high enough that S-wave annihilation to SM states---in the absence of a particle--antiparticle asymmetry---during the CMB epoch is ruled out (see Reference~\cite{Galli:2011rz,Finkbeiner:2011dx,Lin:2011gj,Madhavacheril:2013cna} and discussion in Section~\ref{subsec:CMB}).  The basic reason for the CMB constraint is that the energy released in DM annihilation can distort the surface of last scattering, placing a lower bound on the DM mass.  This can be ameliorated through P-wave annihilation (which occurs if fermionic DM annihilates through a scalar mediator, or if scalar DM annihilates through a vector boson as was proposed for MeV DM; see Reference~\cite{Boehm:2003hm,Boehm:2003ha,Hooper:2007tu}).  In this case, the annihilation rate is suppressed by the velocity of the DM in the late Universe (in the Milky Way, $v \sim 10^{-3}$, while for the smooth background, $v \sim \sqrt{T/m_\chi}$), whereas in the early Universe (at freeze-out, the velocity is $v \sim \frac{1}{3}$) the rate is similar to the S-wave case. 
For the S-wave case, one finds the relic abundance  parameter in Equation~\ref{eq:yRA}, which enters into the direct detection cross section from Equation~\ref{eq:eDD} as follows:
\beq
y_{\rm RA} \simeq  10^{-11} \left(\frac{m_\chi}{10 \mbox{ MeV}}\right)^2\left(\frac{x_f}{20} \right) \left(1-\frac{4 m_\chi^2}{m_{A'}^2}\right)^2.
\label{eq:yfoswave}
\eeq
For the case of P-wave annihilation through a scalar mediator, the couplings must be correspondingly (somewhat) larger to compensate the $v^2$ suppression.  The model space of P-wave annihilating scalar DM is shown as the orange band labeled ``elastic scalar'' in Figure~\ref{fig:DDspacefreezeout}.

\subsection{Thermal Freeze-Out of Asymmetric Dark Matter}
\label{subsec:freezeoutADM}

The DM may also have a particle--antiparticle asymmetry that fixes the dominant component of the relic abundance, similar to what occurs for the baryons in the visible sector (for a review, see Reference~\cite{Kaplan:2009ag}).  However, the particle--antiparticle asymmetry of asymmetric DM becomes visible only if the symmetric abundance is efficiently removed through annihilation in the early Universe.  Stated another way, symmetric freeze-out can be obtained as a limit of the more general case of asymmetric freeze-out.  This is similar to the case of baryons and leptons in the SM, which efficiently annihilate through, for instance, $e^+ e^- \rightarrow \gamma \gamma$ and $p \bar p \rightarrow \pi^+ \pi^-$.  In the SM, these processes are highly efficient, such that positrons, antiprotons, and neutrons are extremely rare in the Universe unless they are produced in astrophysical accelerators that generate cosmic rays.  For asymmetric DM to satisfy bounds on DM annihilation rates at the CMB epoch (discussed below in Section~\ref{subsec:CMB}), the particle--antiparticle annihilation rate must have been large enough to effectively remove the anti-particle, placing a lower bound on the annihilation cross section.  This in turn places a lower bound on the scattering cross section via Equation~\ref{eq:eDD} which also serves as a benchmark for light DM direct detection experiments.   Here we briefly summarize the treatment in Reference~\cite{Lin:2011gj} relevant for our purposes.    

Solving the Boltzmann equation for a general particle--antiparticle asymmetry gives rise to the late-time ratio of particle--antiparticle asymmetries \cite{Graesser:2011wi,Lin:2011gj}:
\begin{equation}
  r_\infty \equiv \frac{Y_-}{Y_+}(\infty) \simeq \frac{Y_-(x_f)}{Y_+(x_f)} \exp \left( \frac{ -\eta_\chi \lambda \sqrt{g_*}}{ x_f^{n+1} (n+1) } \right),
  \label{eq:rinfty}
\end{equation}
where $\eta_\chi \equiv Y_\chi - Y_{\bar \chi}$ and $\lambda \equiv 0.265 M_\mathrm{Pl} m_\chi \sigma_0$.  It turns out that numerically one can set $Y_-(x_f)/Y_+(x_f) \sim 1$ to obtain $ r_\infty$.  The  required annihilation cross section to achieve a particle-symmetric component $r_\infty$ ($r_\infty \ll 1$ corresponds to a large asymmetry) is thus
\beq
\langle \sigma v\rangle \simeq c_f \times 5 \times 10^{-26} \mbox{ cm}^3\mbox{ s}^{-1} \times \ln \left(\frac{1}{r_\infty}\right),
\eeq
where $c_f \equiv \left( \frac{x_f}{20}\right)\left( \frac{4}{\sqrt{g_{*,f}}}\right)$.
One can combine this with the bound on the annihilation cross section from the CMB~\cite{Galli:2011rz,Finkbeiner:2011dx,Lin:2011gj,Madhavacheril:2013cna},
\beq
\langle \sigma v \rangle_{\rm CMB} \lesssim \frac{10^{-27} \mbox{ cm}^3\mbox{ s}^{-1}}{f} \left(\frac{m_\chi}{1 \mbox{ GeV}}\right) \left(\frac{1}{r_\infty}\right),
\label{eq:CMBbound}
\eeq
to obtain an upper bound on $r_\infty$,
\beq
r_\infty \ln \left(\frac{1}{r_\infty}\right) \lesssim \frac{0.02}{f c_f }\left(\frac{m_\chi}{1 \mbox{ GeV}}\right),
\eeq
where $f$ parameterizes an ionizing efficiency ($f \sim 1$ for annihilation to charged leptons but is smaller for annihilation to hadronic states).
This corresponds to a lower bound on the S-wave annihilation $\langle \sigma v \rangle_{\rm CMB} = \langle \sigma v \rangle_V$~\cite{Lin:2011gj}:
\beq
\langle \sigma v \rangle \gtrsim 5 c_f \times 10^{-26} \frac{\mbox{ cm}^3}{\mbox{s}}
\left( \ln\left( 40 f c_f \times \frac{1 \mbox{ GeV}}{m_\chi} \right) + \ln \ln \left(40 f c_f \frac{1 \mbox{ GeV}}{m_\chi} \right)\right).
\label{eq:CMBbound}
\eeq 
This S-wave annihilation cross section is a factor of several larger than what is required to fix the relic abundance in the symmetric S-wave case to achieve a sufficient depletion of the symmetric relic abundance (see figure~2 in Reference~\cite{Lin:2011gj}).  The CMB bound thus gives rise to a scattering cross section for the asymmetric fermion shown in Figure~\ref{fig:DDspacefreezeout},
\beq
\sigma_e \gtrsim 4 \times 10^{-39} \mbox{ cm}^2 \left(\frac{10 \mbox{ MeV}}{m_\chi} \right)^2 \left( \frac{\mu_{e \chi}}{0.5 \mbox{ MeV}}\right)^2 \mbox{ln}\left(\frac{40 \mbox{ GeV}}{m_\chi}\right) \frac{(4 m_\chi^2 - m_M^2)^2}{m_M^4},
\label{eq:ADMbenchmark}
\eeq 
which is derived by combining Equations~\ref{eq:DDreference},~\ref{eq:annxsec}, and~\ref{eq:CMBbound} for the vector mediator.
The upward arrow to the right of the ``asymmetric fermion" label in Figure~\ref{fig:DDspacefreezeout} indicates that the CMB bound demands minimum couplings to remove the symmetric abundance.  An elastic scalar DM line is found nearby in Figure~\ref{fig:DDspacefreezeout} since this case requires similar, but slightly larger, couplings.

\subsection{Freeze-In}

Freeze-in is a process that dominantly occurs at low temperatures when the interactions between the hidden sector and the SM are not strong enough to bring DM into equilibrium with the SM~\cite{Hall:2009bx}.  An initially unpopulated DM state is gradually populated through occasional annihilations of SM states to DM.  For example, if DM is lighter than the pion mass, the dominant freeze-in process is via electrons or plasmons ($\gamma^*$):
\beq
e^+ e^- \rightarrow \chi \bar \chi~~~\text{or}~~~\gamma^* \rightarrow \chi \bar \chi. 
\eeq 
For the purposes of the estimate here, we focus on the first process, though the second one affects the expected scattering rate in direct detection by up to a factor of $10$ for $m_\chi \sim \mbox{keV}$ while having little impact for $m_\chi \sim \mbox{MeV}$~\cite{Dvorkin:2019zdi}.  In the case that the DM is not initially in thermal equilibrium, the Boltzmann equation becomes
\beq
\frac{dY}{dT} = - \frac{\langle \sigma |v| \rangle}{H T s} n^2.
\eeq
If we take $\langle \sigma |v| \rangle \sim \frac{g_\chi^2 g_e^2}{4 \pi T^2}$ as expected for infrared-dominated effects, one finds the following \cite{Chu:2011be,Dvorkin:2019zdi}:
\beq
Y \sim 10^{-4} \frac{g_\chi^2 g_e^2 M_{\rm Pl}}{T}. 
\label{eq:freezeinY}
\eeq
Taking $T \sim 1 \mbox{ MeV}$, where the electrons themselves drop out of thermal equilibrium, and fixing $Y$ by the observed abundance through the relation $\rho_\chi^0 = m_\chi Y s_0$, where $s_0$ is the entropy density today, we obtain the approximate scaling 
\beq
g_\chi g_e \sim 10^{-12} \sqrt{\frac{1 \mbox{ MeV}}{m_\chi}},
\label{eq:freezeingchige}
\eeq  
where the number density of DM $Y$ scales as $1/m_\chi$ in solving for the couplings.
For $m_\chi \lesssim m_e$, the direct detection cross section, using Equation~\ref{eq:DDreference}, then scales as
\beq
\sigma_e \simeq 10^{-39} \mbox{ cm}^2 \left( \frac{m_\chi}{1 \mbox{ keV}}\right).
\label{eq:freezeinbenchmark}
\eeq
This estimate corresponds to the blue band labeled ``freeze-in" in Figure~\ref{fig:DDspacefreezein} for $m_\chi \lesssim 1 \mbox{ MeV}$.  For $m_\chi \gtrsim m_e$, the freeze-in temperature in Equation~\ref{eq:freezeinY} is $T \sim m_\chi$, and the scattering cross section becomes approximately independent of $m_\chi$.  At even higher masses ($m_\chi \gtrsim m_\pi$), new processes involving pions enter; we do not estimate this rate here.  One can see, however, that over much of the mass space, simple estimates allow one to obtain an approximate expected interaction cross section with electrons.

\subsection{Strongly Interacting Massive Particles and Elastically Decoupling Relics}

DM freeze-out can be dominated by $3 \rightarrow 2$ processes~\cite{Hochberg:2014dra,Hochberg:2014kqa} with a ``cross section'' parameterized by
\beq
\langle \sigma_{3 \rightarrow 2} v^2 \rangle \equiv \frac{\alpha^3}{m_\chi^5}.
\eeq  
(Note that this cross section does not carry the usual dimensions, but is defined so that $\Gamma_{3 \rightarrow 2} = n_\chi^2 \langle \sigma_{3 \rightarrow 2} v^2 \rangle$ gives a rate.)  Parametrically $\alpha$ is proportionate to a dark gauge coupling constant $\alpha_D \equiv g_D^2/4\pi$, though here, following References~\cite{Hochberg:2014dra}, ~\cite{Hochberg:2014kqa}, and ~\cite{Kuflik:2017iqs}, $\alpha$ is allowed to absorb ${\cal O}(1)$ factors. 
The observed relic abundance is obtained when
\beq
\alpha \simeq 0.3 \left(\frac{m_\chi}{10 \mbox{ MeV}} \right).
\label{eq:alphaSIMP}
\eeq
These processes continually dump kinetic energy into the DM by cannibalizing the DM's rest energy.  This is observationally ruled out since the DM would be much warmer than observed~\cite{Carlson:1992fn,deLaix:1995vi}.  The solution to this problem, as proposed in References~\cite{Hochberg:2014dra} and~\cite{Hochberg:2014kqa}, is to have a light mediator that bleeds off the energy into the SM sector continuously, typically via a kinetic mixing parameter $\epsilon$ between a dark photon and a visible one.  In this picture, the elastic scattering process that bleeds off the excess energy decouples after the freeze-out of the $3 \rightarrow 2$ process. 

This idea was generalized in References~\cite{Kuflik:2015isi} and~\cite{Kuflik:2017iqs} by allowing $3 \rightarrow 2$ processes in the hidden sector as well as $2 \rightarrow 2$ processes (notably $\chi \bar \chi \rightarrow e^+ e^-$) as in freeze-out to play a role.  Since the  $\chi \bar \chi \leftrightarrow e^+ e^-$ decouples when both directions are in equilibrium, this was called an elastically decoupling relic (ELDER)~\cite{Kuflik:2015isi,Kuflik:2017iqs}.  Thus, the ELDER smoothly interpolates between the elastic freeze-out case (for large enough kinetic mixing, where the $3 \rightarrow 2$ process freezes out well before the annihilation to SM) and the SIMP case (where elastic decoupling with the SM occurs before the $3 \rightarrow 2$ process freezes out).  In the ELDER limit, the relic abundance is set dominantly by the decoupling temperature from the SM, which is given by the following~\cite{Kuflik:2017iqs}
\beq
y_{\rm ELDER} = \epsilon^2 \alpha_D \left(\frac{m_\chi}{m_{A'}} \right)^4 \simeq 6 \times 10^{-15} \left(\frac{g_{*,d}}{10} \right)^{1/2}\left(\frac{m_\chi}{10}\right)\left(\frac{x_d}{17} \right)^6,
\label{eq:yELDER}
\eeq
where $\epsilon$ and $\alpha_D$ are defined above, and $x_d = m_\chi/T_d$ with $T_d$ denoting the elastic decoupling temperature.  This allows one to give a precise benchmark for the ELDER scenario (see the orange band labeled ELDER (SIMP $\uparrow$) in Figure~\ref{fig:DDspacefreezeout}).

\section{Astrophysical, Cosmological, and Accelerator Probes}

When considering whether a DM candidate of a very low mass is detectable in a terrestrial experiment, a wide range of astrophysical, cosmological, and collider constraints must be taken into consideration.  Here we summarize the main features; for further discussion regarding the interplay of the constraints with terrestrial observations, we refer the reader to References~\cite{Lin:2011gj},~\cite{Zurek:2013wia}, and~\cite{Knapen:2017xzo}.


\subsection{Big Bang Nucleosynthesis}

BBN is a powerful constraint if the DM is lighter than an MeV in mass.  Measurements of hydrogen, deuterium, and helium---synthesized dominantly when the Universe had a temperature $T_{\rm BBN} \sim 0.1\text{--}1 \mbox{ MeV}$ at a time $t \sim 1 \mbox{ s}$---indicate that the Universe was expanding at a rate consistent with a relativistic SM photon and three neutrino species.  If any other state were in equilibrium with the SM ({\em e.g.}, through $\chi \bar \chi \leftrightarrow f \bar f$ interactions), this would cause the Universe to expand more quickly and would change the relative abundance of the light elements, parameterized by the effective number of neutrino species, $N_{\rm eff}$.  Current data  constrain $\Delta N_{\rm eff}^{\rm BBN} \lesssim 0.2$~\cite{Yeh:2020mgl,Yeh:2022heq}, implying a more than $3 \sigma$ tension with a single real scalar having a temperature similar to that of neutrinos.  This implies that any DM candidate with a mass $\lesssim 10 \mbox{ MeV}$ has its couplings to the SM constrained by BBN.  

Note that in general, for a DM candidate to be detectable, we must consider not only the DM degrees of freedom but also the mediator that couples the DM to the SM.  The mediator must also either ({\em a}) be heavier than $\sim 1\text{--}10$~MeV, ({\em b}) have sufficiently small couplings to the SM that it remains out of equilibrium, or ({\em c}) contribute $\Delta N_{\rm eff}^\mathrm{BBN} \gtrsim 0.57$ (for a real scalar DM) and be in $\sim 4 \sigma$ tension with BBN.  This limits the types of DM models that can be detected in terrestrial experiments (for an extensive discussion and application to terrestrial experiments, see Reference~\cite{Knapen:2017xzo}; for updated BBN constraints, see References~\cite{Yeh:2020mgl} and~\cite{Yeh:2022heq}). 

For example, BBN constraints are the reason why, for the model space detectable by terrestrial direct detection experiments, DM lighter than an MeV should be mediated by a very light particle.  Direct detection experiments are sensitive to extremely small couplings via light mediators (as in Equation~\ref{eq:freezeingchige}) because of the direct detection enhancement at low momentum transfers (as in Equation~\ref{eq:DDreference} with $m_M \rightarrow 0$).  For such small couplings, the hidden sector is not in equilibrium with the visible sector except at larger couplings (see the shaded orange region in Figure~\ref{fig:DDspacefreezein}).  Likewise, for DM mediated through a heavier particle (having mass $m_M \gtrsim m_\chi v$, where $v$ is the Milky Way virial velocity), the couplings are much larger (see, {\em e.g.}, Equation~\ref{eq:FOxsec} for typical freeze-out couplings).  In this case, the hidden sector generically comes into equilibrium with the visible sector, such that hidden sectors with masses below a few MeV are already ruled out by BBN; this is one reason that the range of the plot in Figure~\ref{fig:DDspacefreezeout} extends only to $m_\chi \gtrsim 1\mbox{ MeV}$.

\subsection{Cosmic Microwave Background}
\label{subsec:CMB}

As discussed above in Sections~\ref{subsec:freezeout} and~\ref{subsec:freezeoutADM}, the CMB places a variety of constraints on a DM candidate of a very low mass.  First, the CMB also places a constraint, at $2 \sigma$, roughly consistent with the BBN bound, $\Delta N_{\rm eff}^{\rm CMB} \lesssim 0.6$.  Since the CMB epoch is at a redshift $z \sim 10^3\text{--}10^4$, while the BBN epoch is at $z \sim 10^{10}$, one cannot simply assume that if the constraint on additional thermal relativistic species is met at one epoch, it is also met at a different epoch.  Note that in the future, CMB Stage IV will dramatically improve on this constraint, $\sigma(N_{\rm eff}^{\rm CMB}) \approx0.03$~\cite{Yeh:2022heq}.

In addition, the CMB is very sensitive to any additional ionizing radiation that can be dumped into the photon--baryon bath.  For example, the Planck limit on DM energy injection due to annihilation is given by Equation~\ref{eq:CMBbound}. 
Since, to set the thermal relic abundance for DM through annihilation, we require $\langle \sigma v \rangle \sim 3 \times 10^{-26} \mbox{ cm}^3\mbox{ s}^{-1}$, this implies that light (sub-GeV) DM cannot have its relic abundance set by S-wave annihilation while remaining consistent with CMB-epoch ionization constraints.  There are two ways to circumvent this constraint, as discussed in Section~\ref{subsec:freezeout}: P-wave annihilation where the cross section is $v$ suppressed, or asymmetric DM where $r_\infty \ll 1$.  As highlighted above, models of asymmetric DM must have an efficient mechanism for removing the symmetric abundance of DM early in the Universe.  This implies a lower bound on the annihilation cross section in the early Universe (shown in Figure~\ref{fig:DDspacefreezeout} as the asymmetric fermion curve) to adequately dilute the symmetric abundance.  


\subsection{Large-Scale Structure}

Large-scale structure places two important bounds on DM having a very low mass.  The first important bound is the warm DM bound, from the formation of structure on small scales, as observed with Lyman-$\alpha$ forest and other large-scale structure measurements of DM clustering~\cite{DES:2020fxi,Enzi:2020ieg}.  Thermalized DM has the following relation between its velocity and temperature:
\beq
v = \sqrt{\frac{3 T}{m_\chi}}.
\eeq
If DM has its temperature on the same order as the visible temperature $T \sim 10^{-4} \mbox{ eV}$, and we require DM to be cold enough for it to fall into structures having virial velocity $v \sim 10^{-4}$ ({\em e.g.}, as observed in dwarf galaxies), this implies that DM should have the following mass:
\beq
m_\chi \gtrsim 10 \mbox{ keV}.
\eeq 
This is the back-of-the-envelope estimate of the warm DM bound (which can be more carefully derived by simulating the formation of structures with warm DM; see, {\em e.g.}, References~\cite{DES:2020fxi} and~\cite{Enzi:2020ieg}).  DM may still be lighter than this bound, but its temperature should not be set by the interactions with the SM ({\em e.g.}, as happens with non-thermal mechanisms such as the misalignment mechanism for the axion or inflationary production of vector DM)~\cite{Nelson:2011sf,Graham:2015rva}.

If DM has an interaction with the SM through a relatively light gauge boson, this can also cause late kinetic decoupling between the SM and the dark sector and modify the matter power spectrum on small scales.  For example, DM can couple to neutrinos through a mediator (product of couplings to the mediator $g_\chi g_\nu$), giving rise to a kinetic decoupling temperature:
\beq
T_{\rm kd} \simeq \mbox{keV} \frac{m_M}{\mbox{MeV}} \left(\frac{m_\chi}{\mbox{MeV}}\right)^{1/4}\left(\frac{10^{-6}}{g_\chi g_\nu}\right)^{1/2}.
\eeq
Such a kinetic decoupling temperature is bounded by the Lyman-$\alpha$ forest (see Reference~\cite{Hooper:2007tu}).

DM of a very low mass also tends to come with new light forces that can mediate large self-interactions.  The self-interaction rate is bounded by the shape of DM halos, which implies that DM self-interactions should be rare enough that the formation of halos is dominated by gravitational interactions.  This implies the following~\cite{Tulin:2017ara}:
\beq
\frac{\sigma}{m_\chi} \lesssim 0.1 \text{--} 10 \mbox{ cm}^2\mbox{ g}^{-1},
\eeq
where the large range of possible bounded cross sections is due to the large range of system mass scales and interaction times (from dwarf galaxies at $10^{7}~M_\odot$ to clusters of galaxies at $10^{15}~M_\odot$) as well as the still-somewhat-imprecise nature of simulations, which depends on baryonic effects and their (non-linear) feedback on the DM structure formation process.  Especially when the mediator is light, this places severe limits on the DM--mediator coupling (taking $\frac{\sigma}{m_\chi} = 1 \mbox{ cm}^2\mbox{ g}^{-1}$):
\beq
\alpha_\chi \lesssim 6 \times 10^{-10} \times \left(\frac{m_\chi}{1 \mbox{ MeV}}\right)^{3/2}.
\eeq
When the mediator becomes massive ($m_\phi > m_\chi v$), we have a less severe limit:
\beq
\alpha_\chi \lesssim 0.025 \left(\frac{1 \mbox{ keV}}{m_\chi}\right)^{1/2} \left(\frac{m_\phi}{1 \mbox{ MeV}}\right)^2.
\eeq
For further discussion, we refer the reader to References~\cite{Hochberg:2015fth} and~\cite{Knapen:2017ekk}.

\subsection{Stellar Evolution and Cooling}

\begin{figure}
\includegraphics[width=0.45\linewidth]{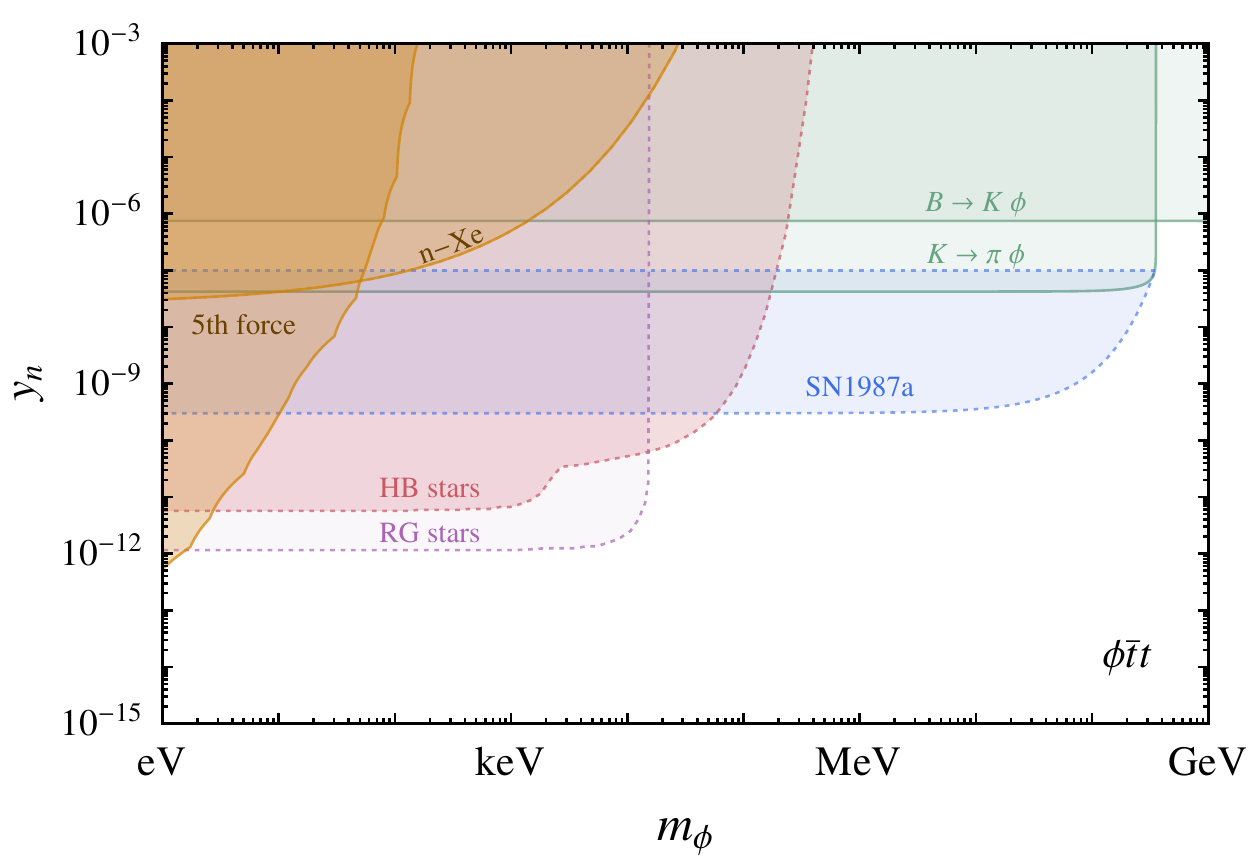}
\includegraphics[width=0.45\linewidth]{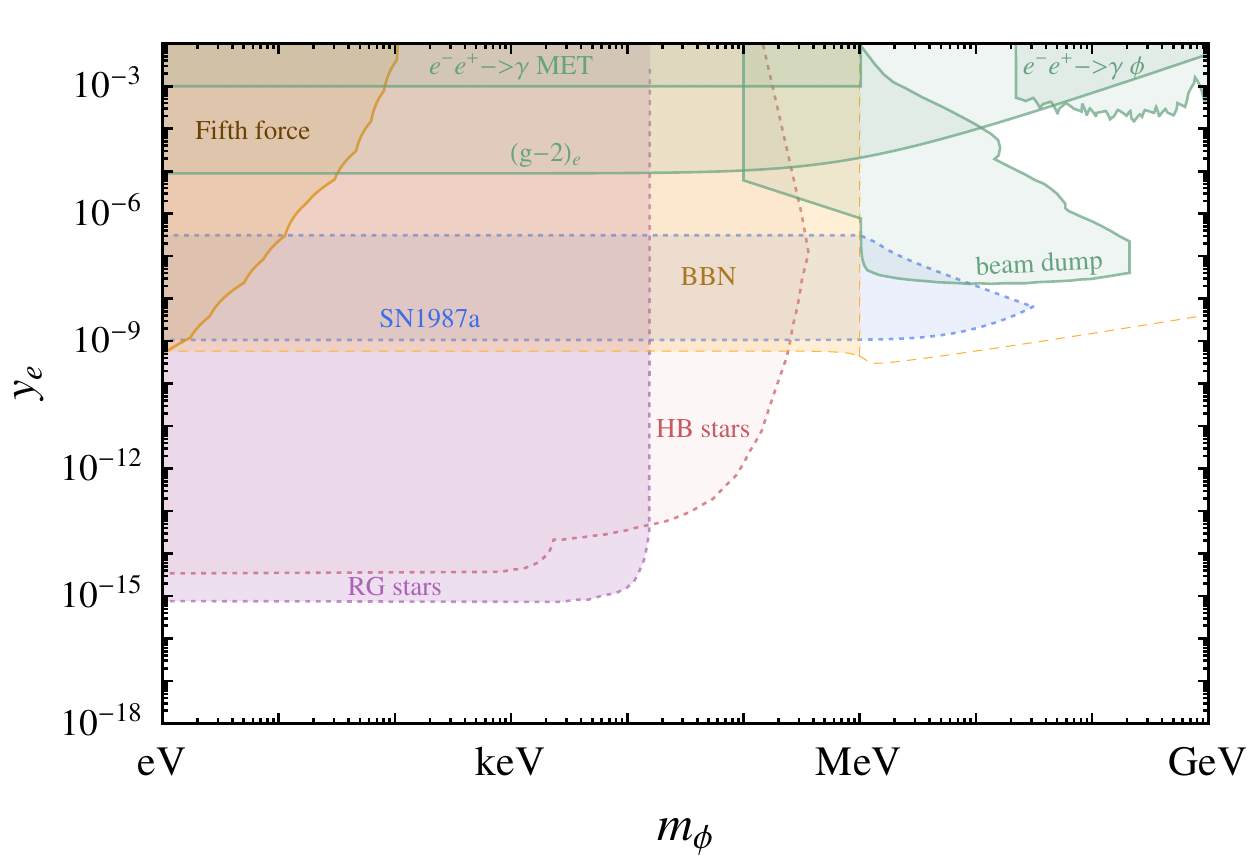}
\caption{Constraints on ({\em a}) mediator-nucleon coupling ${\cal L} = y_n \phi \bar n n$ (mediated by a top quark) and ({\em b}) mediator-electron coupling ${\cal L} = y_e \bar e e$ as a function of the mediator mass. Abbreviations: BBN, big bang nucleosynthesis; HB, horizontal branch; MET, missing transverse energy; RG, red giant. Figure adapted from Reference~\cite{Knapen:2017xzo}.\label{fig:mediatorspace}
}
\end{figure}

When the DM, or the mediator, is lighter than the temperature of a star, production of dark-sector particles can lead to cooling that is more efficient than what happens in the SM, where states are strongly coupled to the stellar plasma (for a classic review, see Reference~\cite{Raffelt:1996wa}).  This cooling changes the evolution of the star from that predicted by standard theory, and it could be observed.  This has long been appreciated in particular for axions, but it also becomes relevant for DM and mediators with masses $\lesssim 30 \mbox{ MeV}$, the temperature of a supernova.  In addition to supernovae, red giants (RGs; relevant for masses $\lesssim 10 \mbox{ keV}$), horizontal branch stars (HBs; $\lesssim 100 \mbox{ keV}$) for nucleon couplings, and white dwarves (WDs; $\lesssim 10 \mbox{ keV}$) are all relevant.  Constraints from these stellar cooling limits are shown as shaded regions in Figure~\ref{fig:DDspacefreezein} and~\ref{fig:mediatorspace}.  When the coupling becomes large enough, particularly in dense environments like the neutron stars that form inside supernovae, trapping becomes relevant.  The supernova cooling bounds have recently been updated in Reference~\cite{Chang:2018rso}, while the RG, HB, and WD stellar cooling bounds for mediator couplings to nucleons, electrons, and dark photons are quoted in detail in Reference~\cite{Knapen:2017xzo}. 

Light DM, especially if it carries a particle--antiparticle asymmetry that prevents annihilation, can accumulate in the centers of stars and also cause a change in their evolution.  This has been considered in particular for neutron stars, either due to inducing an instability (for a summary, see the discussion in Reference~\cite{Zurek:2013wia}) or due to kinetic heating~\cite{Baryakhtar:2017dbj}.  Accumulation of light DM can also lead to modification of main branch stellar evolution~\cite{Taoso:2010tg} and in brown dwarves~\cite{Zentner:2011wx} and WDs~\cite{Leung:2013pra,Graham:2018efk}. For a fairly up-to-date list of references regarding compact star constraints on HSDM interactions with electrons and nucleons, we refer the reader to Reference~\cite{Joglekar:2020liw}.


\subsection{Accelerator Probes of Light Hidden Sectors}

Intensity experiments, such as beam dumps featuring a large number of protons on target, can probe relatively light (typically sub-GeV mass) hidden-sector particles.  The constraints are most direct on the mediating particle (vector or scalar) and depend on the decay channel, whether to visible (notably $e^+ e^-$) or invisible particles.

\begin{itemize}

\item Invisible decays:  In the case of invisible decays, the mediator either decays dominantly to DM, or is stable on detector timescales.  If the mediator couples to hadrons and is lighter than $\Lambda_{\rm QCD}$, it can be constrained through the invisible processes $B \rightarrow K \phi$ and $K \rightarrow \pi \phi$, where $\phi$ is a scalar or vector mediator~\cite{Knapen:2017xzo}.  The effective coupling to nucleons constrained in this way is $y_n \lesssim 10^{-5}\text{--}10^{-7}$, depending on the flavor structure of the $\phi$--quark coupling.  If the mediator couples to electrons, it can be constrained by $e^+ e^- \rightarrow \gamma \phi$, that is, a monophoton search in BaBar.  The electron $(g-2)$ measurement also provides a powerful bound. For a summary of collider bounds, we refer the reader to References~\cite{Izaguirre:2013uxa},~\cite{Izaguirre:2014bca}, and~\cite{Knapen:2017xzo}.
\item Visible decays: For visible decays, the constraints are strongest when the mediator is produced, and decays directly to charged leptons $\ell^+ \ell^-$~\cite{Bjorken:2009mm,Andreas:2012mt}.  These constraints are typically labeled as beam-dump constraints.

\end{itemize}

While we have not exhaustively summarized the accelerator probes on light hidden sectors, these summarize the general types of constraints on mediators.  We show two summary plots in Figure~\ref{fig:mediatorspace} for nucleon and lepton couplings that demonstrate the collider, stellar, and fifth-force bounds on the nucleon--scalar and electron--scalar couplings.

\section{Direct detection of light-particle dark matter}

The direct detection of light DM depends on considerations of both kinematics and interactions.  That is, one needs to find a target material with a strong response (i.e., Dynamic Structure Factor) for energy and momentum deposition in the DM kinematic regime for the relevant type of interaction.  We briefly summarize both before describing the particular mechanisms for detection of light DM.

\begin{figure}
\includegraphics[width=0.45\linewidth]{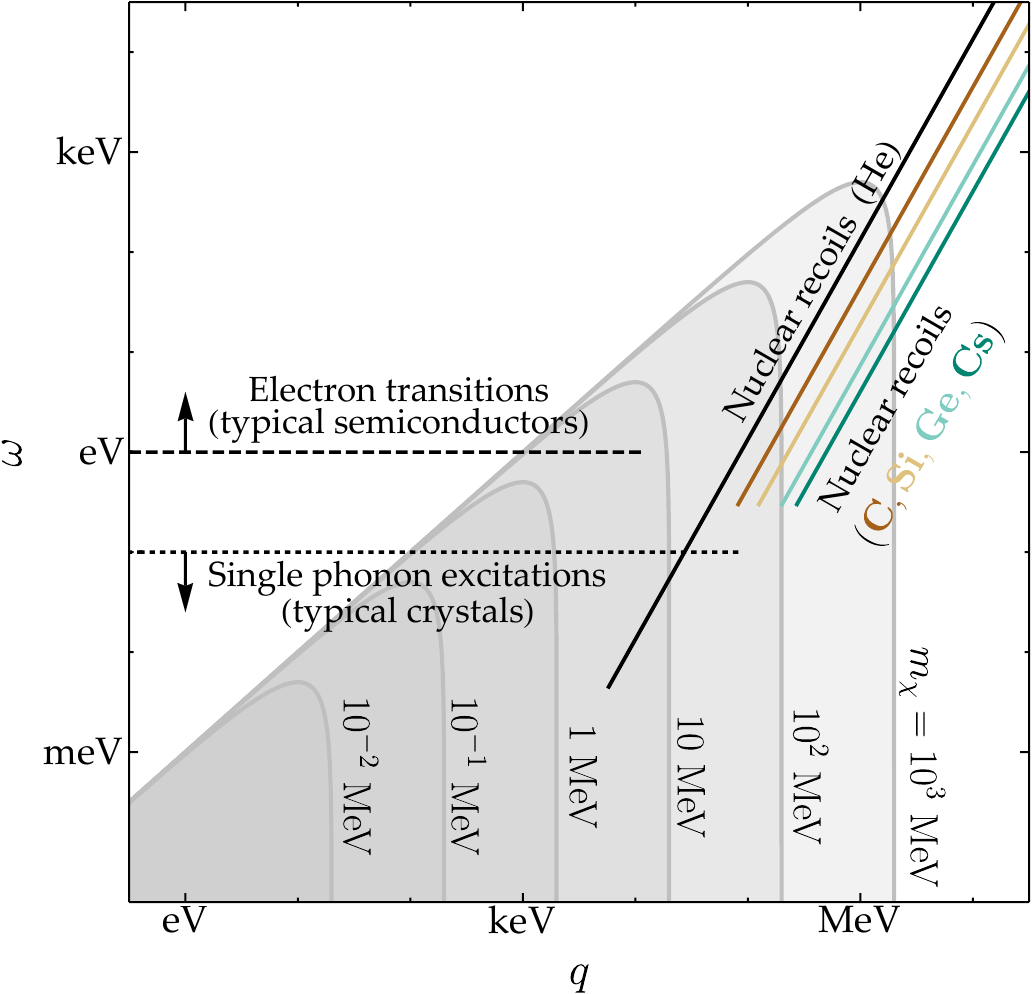}
\caption{The kinematic parameter space of light dark matter in terms of the energy deposition ($\omega$) and momentum transfer ($q$) in a direct detection experiment.  The shaded parabolas correspond to the range of energy and momentum that a dark matter particle of a given mass can deposit (Equation~\ref{eq:parabola}).  The nuclear recoil curves correspond to the energy deposit on a nucleus (Equation~\ref{eq:NR}).  The horizontal lines at $1$ eV and $100$ meV correspond to the energy gaps (as $q \rightarrow 0$) of electronic excitations (in semiconductors) and optical phonons (in crystals).  Figure adapted from Reference~\cite{Trickle:2019nya}.
\label{fig:kinematics}
}
\end{figure}

In direct detection, DM must be able to cause a transition from the initial state to the final state $|i \rangle \rightarrow |f \rangle$ of the target system, with the DM typically depositing some momentum ${\bf q} = {\bf p} - {\bf p}'$ (where ${\bf p } = m_X {\bf v}$ is the initial DM momentum and ${\bf p}'$ is the final DM momentum), which corresponds to some energy $\omega$ deposited by the DM on the target:
\beq
\omega = \frac{1}{2} m_\chi v^2 - \frac{(m_\chi {\bf v} - {\bf q})^2}{2 m_\chi} = {\bf q} \cdot {\bf v} - \frac{q^2}{2 m_\chi}.
\eeq
Note that this expression assumes that the DM has no internal excitations (such as energy levels) that could cause the transition to be inelastic on the DM side.
The energy deposited is bounded by a parabola, 
\beq
\omega \leq q v_{\rm max} - q^2/2 m_\chi,
\label{eq:parabola}
\eeq 
as first illustrated in Reference~\cite{Trickle:2019nya} and reproduced in Figure~\ref{fig:kinematics}.    This figure shows the basic kinematic requirements of a material to allow for a detection event: The target material must have a state whose transition energy and momentum lie below (and within) the DM kinematic parabola.  
In addition, the closer a mode in the material is to the upper part of the parabola, the more energy can be read out, allowing in many cases for a more viable path toward detection.  
DM can never transfer more momentum than the following: 
\beq
q_\mathrm{BW} = 2 m_\chi v_{\rm max},
\eeq 
which represents the ``brick wall'' limit at which little energy is deposited, corresponding to the right edge of the DM kinematic parabola.  Take, for example, the case of nuclear recoils, where the energy deposit is \beq
 \omega = \frac{q^2}{2 m_N},
 \label{eq:NR}
 \eeq 
 where $m_N$ is the target nucleus mass, also shown in Figure~\ref{fig:kinematics}.  At the point where the nuclear recoil kinematic curve intersects the DM parabola, only a small fraction of the DM kinetic energy is deposited onto the nucleus, 
\beq
\omega = \frac{2 m_\chi^2 v^2}{m_N} \ll \frac{m_\chi v^2}{2},
\eeq 
when the DM is lighter than the typical SM nucleus mass.  One immediately sees that sub-GeV DM is kinematically poorly matched to SM nuclei, and one should search for other targets for DM interactions.  We now summarize the types of excitations that are relevant in each mass regime.   

\subsection{Targets and Excitations for Dark Matter Interactions}


\begin{itemize}
\item For DM with mass  $\gtrsim 30$ MeV, nuclear recoils are a relevant target, though they extract only a small fraction of the DM kinetic energy. Since the DM energy deposition is given in Equation~\ref{eq:NR}, in order to extract the maximum energy deposition $\omega$, the lightest target is preferable. In particular, superfluid helium has been identified as a promising target for the TESSERACT DM experiment.  The lower end of the DM mass range $m_\chi \gtrsim 30$~MeV corresponds to a DM energy deposit on nuclei (via Equation~\ref{eq:NR}) larger than $\omega \gtrsim 500$~meV, corresponding to the typical energy of collective excitations, where the nucleus can no longer be treated as free (see discussion in Reference~\cite{Trickle:2019nya}).  For smaller energy depositions, the relevant excitations are phonons.  

\item For DM with mass $m_\chi \gtrsim 1$~MeV and kinetic energy $\omega \gtrsim 1 \mbox{ eV}$, electrons in targets such as semiconductors (having a typical band gap of $1$~eV) and ionization in noble liquids such as xenon (with an ionization threshold of $\sim 10~$eV) are good targets and have been explored extensively~\cite{Essig:2011nj,Graham:2012su,Essig:2015cda}.  Ripping an electron out of a two-dimensional material such as graphene, with a work function also in the $1$-eV energy range~\cite{Hochberg:2017wce} and breaking chemical bonds~\cite{Essig:2016crl} with energy in the tens-of-eV range, has also been proposed.

\item At slightly lower energy depositions, in the $100$-meV range, collective excitations such as optical phonons (gapped excitations having energy $\omega_{\rm ph}$) become available and have been proposed as a viable pathway~\cite{Knapen:2017ekk,Griffin:2019mvc}, having greatest sensitivity via scattering for DM in the  $m_\chi \gtrsim $~keV--MeV mass range.  Collective excitations are highly sensitive to heavier DM as long as the momentum transfer is smaller than that required to kick the ion out of the lattice potential $q \lesssim \sqrt{2 m_N \omega_{\rm ph}} \sim 10\text{--}100 \mbox{ keV}$.  This momentum corresponds to that carried by MeV mass DM in the Milky Way, though for heavier DM only a small fraction of its total momentum may be transferred, especially if the mediator is light, having a $1/q^4$ enhancement in the cross section as in Equation~\ref{eq:DDreference} with $m_M \rightarrow 0$ (for a detailed analysis, see Reference~\cite{Griffin:2019mvc}). The SPICE (sub-eV Polar Interactions Cryogenic Experiment) experiment has research underway to detect single phonons in crystals~\cite{Anthony-Petersen:2022ujw}.   In addition to phonons, magnons are a different type of collective excitation in materials that can detect spin-dependent DM interactions~\cite{Trickle:2019ovy}.

\item Other gapped excitations, such as vibrational degrees of freedom in organics~\cite{Blanco:2019lrf}, and electronic excitations with small gaps, such as occur in superconductors~\cite{Hochberg:2015pha,Hochberg:2015fth}, Dirac materials~\cite{Hochberg:2017wce,Chen:2022pyd}, heavy fermion materials~\cite{Hochberg:2021pkt}, and doped semiconductors~\cite{Du:2022dxf}, are also viable targets for $m_\chi \gtrsim$~keV DM.   
\end{itemize}

So far, we have explored the importance of kinematic matching between DM and target.  We now examine the strength of the target response to a given energy $\omega$ and momentum ${\bf q}$ deposition.  This requires computing the quantum mechanical matrix element entering into Fermi's Golden Rule for interaction rates.  The matrix element in turn depends on the interaction type, such as spin-independent, spin-dependent, or a more generic interaction such as an electric or magnetic dipole or anapole.  In the next sub-sections, we lay out the general quantum mechanical framework for computing the rate and then apply it to spin-independent scattering.  In Section~\ref{sec:EFT}, we consider more general interactions in an effective field theory (EFT) framework.

\subsection{Quantum Mechanics of Dark Matter Scattering}
\label{subsec:QM}

The interaction type enters directly into the calculation of the matrix element for the DM to induce a transition from an initial to a final state in a target material.  The DM deposits some energy and momentum $(\omega,{\bf q})$ within the kinematically allowed DM parabola shown in Figure~\ref{fig:kinematics}, and the over-arching goal is to find a material with a strong quantum mechanical response, encapsulated in a Dynamic Structure Factor.   For a more complete discussion of the physics reviewed here, we refer the reader to References~\cite{Trickle:2019nya} and~\cite{Trickle:2020oki}.  


Quantum mechanics enters in determining the transition rate from the initial state to the final state of the target material $|i \rangle \rightarrow |f \rangle$ via
a simple application of Fermi's Golden Rule:
\beq
\Gamma({\bf v}) = V \int \frac{d^3 q}{(2 \pi)^3} \sum_f |\langle {\bf p}',f|H_{\rm int} | {\bf p},i \rangle|^2 2 \pi \delta(E_f - E_i - \omega), 
\label{eq:FGR}
\eeq
where $H_{\rm int}$ is the interaction Hamiltonian, and we assume throughout that the DM state factorizes from the target state (since they are unentangled), $|{\bf p},i \rangle = |{\bf p} \rangle \otimes | i \rangle$, {\em etc.}, and $V$ is the volume of the target.     The interaction rate in Equation~\ref{eq:FGR} depends on the DM velocity ${\bf v}$, whose typical value in the Milky Way galaxy is $v \sim 10^{-3}$.  The detectable interaction rate, per unit of target mass, is extracted by integrating over the DM velocity phase space:
\beq
R = \frac{1}{\rho_T} \frac{\rho_\chi}{m_\chi} \int d^3 v f_\chi({\bf v}) \Gamma({\bf v}),
\label{eq:integratedrate}
\eeq
where $\rho_{\chi}$ and $\rho_{T}$ are the DM and target densities, and $f_\chi$ is typically taken to be a Maxwell--Boltzmann distribution truncated at the escape velocity $v_{\rm esc} \sim 600 \mbox{ km}\mbox{ s}^{-1}$ of the DM from the galaxy. Note that while $\Gamma({\bf v})$ is a rate, $R$ is a rate per unit of mass.


To illustrate the principles discussed above, we first consider DM scattering via a spin-independent interaction.
The DM creates a potential,
\beq
\langle {\bf p}' | H_{\rm int} | {\bf p} \rangle =  V({\bf q})
\eeq
to which the target responds.  It is convenient to factorize this potential into a material response ${\cal F}_T$ (which is agnostic about the DM) and a DM matrix element ${\cal M}$ (which is agnostic about the target):
\beq
V({\bf q}) = {\cal M}(q) {\cal F}_T({\bf q}).
\label{eq:potential}
\eeq
For spin-independent scattering, the DM-induced potential ${\cal M}(q)$ is characterized by a fiducial cross section and a mediator form factor:
\beq
{\cal M}(q) = {\cal M}(q_0) {\cal F}_{\rm med}(q),
\eeq
where ${\cal F}_{\rm med}(q)$ is either 1 (for a heavy mediator) or $(q_0/q)^2$ (for a light mediator with momentum dependence typical of Rutherford scattering).  ${\cal M}(q_0)$ is related to a reference cross section convenient for parameterizing the overall strength of the interaction via
\beq
\bar \sigma_T \equiv \frac{\mu_{\chi T}^2}{\pi} |{\cal M}_{\chi T}(q_0)|^2_{q_0 = m_\chi v_0},
\label{eq:sigmaconvention}
\eeq 
where $\mu_{\chi T}$ ($T = n$ or $e$) denotes the DM--target (nucleon or electron) reduced mass.
The target response ${\cal F}_T(q)$, for spin-independent interactions, is traditionally absorbed into what is known as the Dynamic Structure Factor ($S$), which characterizes the response of the material:
\beq
S({\bf q},\omega) \equiv \frac{1}{V} \sum_f |\langle f | {\cal F}_T({\bf q})|i \rangle |^2 2 \pi \delta(E_f - E_i -\omega),
\label{eq:dynamic}
\eeq
where $E_f$ and $E_i$ are the final and initial energies of the target.
The formalism where the DM scattering rate is proportional to the Dynamic Structure Factor is valid as long as the scattering is spin-independent.   Below, we will generalize the calculation to a generic type of potential, including spins; however, for intuition, and to match to the standard condensed matter understanding, we start with the spin-independent case.  
The DM--target scattering rate can thus be written in a unified way in terms of the Dynamic Structure Factor:
\beq
\Gamma({\bf v}) = \frac{\pi \bar \sigma}{\mu^2} \int \frac{d^3 q}{(2 \pi)^3} {\cal F}_{\rm med}^2(q) S({\bf q},\omega).
\label{eq:rate}
\eeq

The three basic types of spin-independent interactions considered---nuclear recoil, electron excitation, and phonon excitation---can be expressed in terms of the Dynamic Structure Factor.  As shown below, we can reproduce the results in the literature using this simple unifying language.  The Dynamic Structure Factor is a material-specific response that depends kinematically only on the input $({\bf q},\omega)$ provided by the DM interaction.  The Dynamic Structure Factor can be generalized from spin-independent interactions within an EFT framework but using the same basic tools introduced here.  We take on that task in the next section, but first we summarize the broad types of interactions used for light DM detection---nuclear recoil, electronic excitations, and single phonon excitation---in the language of the Dynamic Structure Factor.


\subsubsection{Nuclear recoils}  

In the case of nuclear recoils, for each species of nucleus, the Dynamic Structure Factor is
 \beq
 S({\bf q},\omega) = 2 \pi \frac{\rho_T}{m_N} \frac{f_N^2}{f_n^2} F_N^2(q) \delta\left(\frac{q^2}{2 m_N} - \omega \right),
 \label{eq:nuclearrecoilDSF}
 \eeq
 where $m_N$ is the target nucleus mass, $F_N(q)$ is the nuclear form factor (often taken to be the Helm form factor), $f_N$ is the coupling to the nucleus, and $f_n$ is the coupling to a nucleon which we divide through by convention to cancel the same coupling in the cross section (Equation~\ref{eq:sigmaconvention}).  Note that a single nucleus response is appropriate as long as the nucleus can be treated as free.  This occurs when the energy deposition is greater than a typical phonon energy, $\omega \gg \omega_{\rm ph} \simeq 10$--$500$~meV (or, equivalently, $q \gg \sqrt{m_N \omega_{\rm ph}}$).
 
\subsubsection{Electronic excitations} 

At the next level of complexity and at lower energy depositions and momentum transfer, DM interactions can induce electronic transitions.  
In this case, the Dynamic Structure Factor depends on the electron wavefunctions in the initial and final states:
 \beq
 S({\bf q},\omega) = \frac{2 \pi}{V} \left(\frac{f_e}{f_e^0}\right)^2 \sum_{i,f} \delta(E_f - E_i - \omega) \left| \int \frac{d^3 k'}{(2\pi)^3} \frac{d^3 k}{(2\pi)^3} (2 \pi)^3 \delta^3({\bf k}' - {\bf k} - {\bf q}) \psi_f^*({\bf k}') \psi_i({\bf k}) \right|^2, \nonumber
 \eeq
 where the sum over $i$ and $f$ is over all the initial and final electronic states whose energy difference is $\omega = E_f - E_i$. 
  In a semiconductor, the relevant states are core, valence, conduction, and free electrons, where the wavefunctions are written in terms of Bloch waves labeled by a band index $I$ and wavenumber ${\bf k}_1$:
 \beq
 \psi_{I,{\bf k}_1}({\bf x}) & = & \frac{1}{\sqrt{V}} \sum_{{\bf G}_1} u_I({\bf k}_1 + {\bf G}_1) e^{i({\bf k}_1 + {\bf G}_1) \cdot {\bf x}}, \\
 \psi_{I,{\bf k}_1}({\bf k}) & = & \int d^3 x \psi_{I,{\bf k}_1}({\bf x}) e^{-i {\bf k} \cdot {\bf x}} = \frac{1}{\sqrt{V}} \sum_{{\bf G}_1} u_I({\bf k}_1+{\bf G}_1) (2 \pi)^3 \delta^3({\bf k}_1 + {\bf G}_1-{\bf k})
 \eeq
 where ${\bf G}_1$ is a reciprocal lattice vector (Umklapp).
 This immediately leads to a Dynamic Structure Factor, 
 \begin{eqnarray}
 S({\bf q},\omega) &=&  \frac{2}{V} \sum_{i,f} \int_{1{\rm BZ}} \frac{d^3 k_1}{(2 \pi)^3}\frac{d^3 k_2}{(2 \pi)^3} 2 \pi \delta(E_{f,{\bf k}_2} - E_{i,{\bf k}_1} - \omega )  \nonumber \\
 & \times& \left| \sum_{{\bf G}_1,{\bf G}_2} (2 \pi)^3 \delta^3({\bf k}_2 + {\bf G}_2 - {\bf k}_1 - {\bf G}_1 - {\bf q}) u_{f}^*({\bf k}_2 + {\bf G}_2) u_i({\bf k}_1 + {\bf G}_1) \right|^2,
 \end{eqnarray}
 where the prefactor of 2 comes from summing over degenerate spins.  Now we can define a crystal form factor with an Umklapp ${\bf G}$, which lets us take into account  in momentum space the periodicity of the crystal lattice,
 \beq
 f_{\left[i,{\bf k}_1,f,{\bf k}_2,{\bf G}\right]} \equiv \sum_{{\bf G}_1,{\bf G}_2} u_f^*({\bf k}_2 + {\bf G}_2) u_i({\bf k}_1 + {\bf G}_1) \delta_{{\bf G}_2 - {\bf G}_1,{\bf G}},
 \eeq  
so that we can rewrite the Dynamic Structure Factor as follows:
 \begin{eqnarray}
 S({\bf q},\omega) &=&  \frac{2}{V} \sum_{i,f} \int_{1{\rm BZ}} \frac{d^3 k_1}{(2 \pi)^3}\frac{d^3 k_1}{(2 \pi)^3} 2 \pi \delta(E_{f,{\bf k}_2} - E_{i,{\bf k}_1} - \omega )  \nonumber \\
 & \times&\sum_{{\bf G}} (2 \pi)^3 \delta^3({\bf k}_2 - {\bf k}_1 - {\bf G} - {\bf q})  \left| f_{\left[i,{\bf k}_1,f,{\bf k}_2,{\bf G}\right]} \right|^2.
 \end{eqnarray}

This formula can also be applied to the case of superconductors, where the calculation is relatively simple because, for energy depositions well above the Cooper pair binding energy, the electrons behave as free particles in a Fermi-degenerate sea.   In that case, ${\bf G} = 0$ and the crystal form factor is simply replaced by the Fermi--Dirac distributions,
\beq
\left| f_{\left[i,{\bf k}_1,f,{\bf k}_2,{\bf G}\right]} \right|^2 =  f(\omega_{k_1}) (1-f(\omega_{k_2})),
\eeq
where $f(E_i) = \left[1+\mbox{exp}\left(\frac{E_i - \mu_i}{T} \right) \right]^{-1}$ is the Fermi--Dirac distribution of electrons at temperature $T$.  In the limit $T \rightarrow 0$, the Dynamic Structure Factor reduces to the following~\cite{Hochberg:2015fth}:
\beq
S(\omega,{\bf q}) \simeq \frac{m_e^2 \omega}{\pi q}\Theta(q v_F - \omega),
\label{eq:superconductorDSF}
\eeq
where $v_F \simeq 10^{-2}$ is the Fermi velocity of electrons in a superconductor.

 
The technical obstruction to computing DM interaction rates in materials like semiconductors is knowledge of the electronic wavefunctions. In some cases, especially where the electrons are more tightly bound~\cite{Griffin:2021znd}, analytic atomic wavefunctions can provide an approximation that gives correct order-of-magnitude estimates for DM interaction rates~\cite{Lee:2015qva,Hochberg:2016ntt}.  However, especially for conduction electrons, these approximations are not very good.  There now exist multiple codes to compute spin-independent scattering rates of light DM electrons using wavefunctions computed via density functional theory (DFT).  This includes the QED\textsc{ark}~\cite{Essig:2015cda},  QCD\textsc{ark}~\cite{Dreyer:2023ovn}, EXCEED-DM~\cite{Griffin:2021znd,Trickle:2022fwt}, and D\textsc{ark}ELF~\cite{Knapen:2021bwg} packages.  EXCEED-DM extended QED\textsc{ark} by including the all-electron reconstructed wavefunctions and additional electronic states outside of valence and conduction bands.  D\textsc{ark}ELF makes use of the DFT-computed dielectric function $\epsilon(\omega,{\bf q})$ with the observation that the dynamic structure function, {\em for dark photon--like scattering on electrons}, can be written in the low-temperature limit as follows~\cite{Knapen:2021run,Hochberg:2021pkt}:
 \beq
 S(\omega,{\bf q}) = \frac{q^2}{2 \pi \alpha_{em}} \mbox{Im} \left[ \frac{-1}{\epsilon_L(\omega,{\bf q})}\right].
 \eeq
 This can be a convenient expression for spin-independent scattering if the dielectric function is available for all $(\omega,{\bf q})$ of interest. 
 
Table~\ref{tab:ElScatt} summarizes the types of targets, and their gaps, proposed for DM interacting with the electron.  There are already many experiments in progress that realize these ideas:
\begin{itemize}
\item Ionization in atoms and excitation across the band gap in semiconductors were the first processes proposed to detect MeV--GeV DM and are currently being implemented in semiconductor targets in the SuperCDMS, SENSEI, DAMIC, and EDELWEISS experiments.  The Xenon (PandaX, Xenon1T, LUX) and liquid argon (DarkSide) experiments have also done searches for ionization of electrons.  
\item Cooper pair breaking in superconductors has been implemented to search for keV--GeV light DM with superconducting nanowires~\cite{Hochberg:2021yud}.  
\end{itemize}
For a currently complete discussion of the ongoing experimental efforts, which we do not attempt to cite in detail here, we refer the reader to Reference~\cite{Essig:2022dfa} (especially Figure 1).

\begin{table}
\begin{tabular}{c|c|c|c|c}
\hline
Target  & Reaction processes & Typical gap & Elastic or inelastic? & DM Mass Range \\ \hline \hline
Atom & Ionization & 10 eV & Inelastic & $\gtrsim 10$~MeV--GeV \\ \hline
Semiconductor & Excitation across band gap & $\sim 1 \mbox{ eV}$ & Inelastic & MeV--GeV\\ \hline
Superconductor & Cooper pair breaking & $\sim 1 \mbox{ meV}$ & Approximately elastic\footnote{The superconductor DM detection process is elastic for energy depositions well above the Cooper pair binding energy.} & $\gtrsim 1$~keV--GeV\\ \hline
Graphene & Electron ejection & $ \sim 1 \mbox{ eV}$ & Inelastic & $\gtrsim 1$~MeV--GeV\\ \hline
Dirac material &  Excitation across band gap & $\sim 0$--$1 \mbox{ meV}$ & Inelastic & keV--GeV\\ \hline
Heavy fermion material &  Excitation across band gap & $\sim 10 \mbox{ meV}$ & Inelastic & 10~keV--GeV\\ \hline
\hline
\end{tabular}
\caption{Summary of the target materials that have been proposed for dark matter (DM) detection through electron excitation.}\label{tab:ElScatt}
\end{table}

\subsubsection{Single Phonon excitations}  

\begin{figure}
\begin{center}
\includegraphics[width=0.4\textwidth]{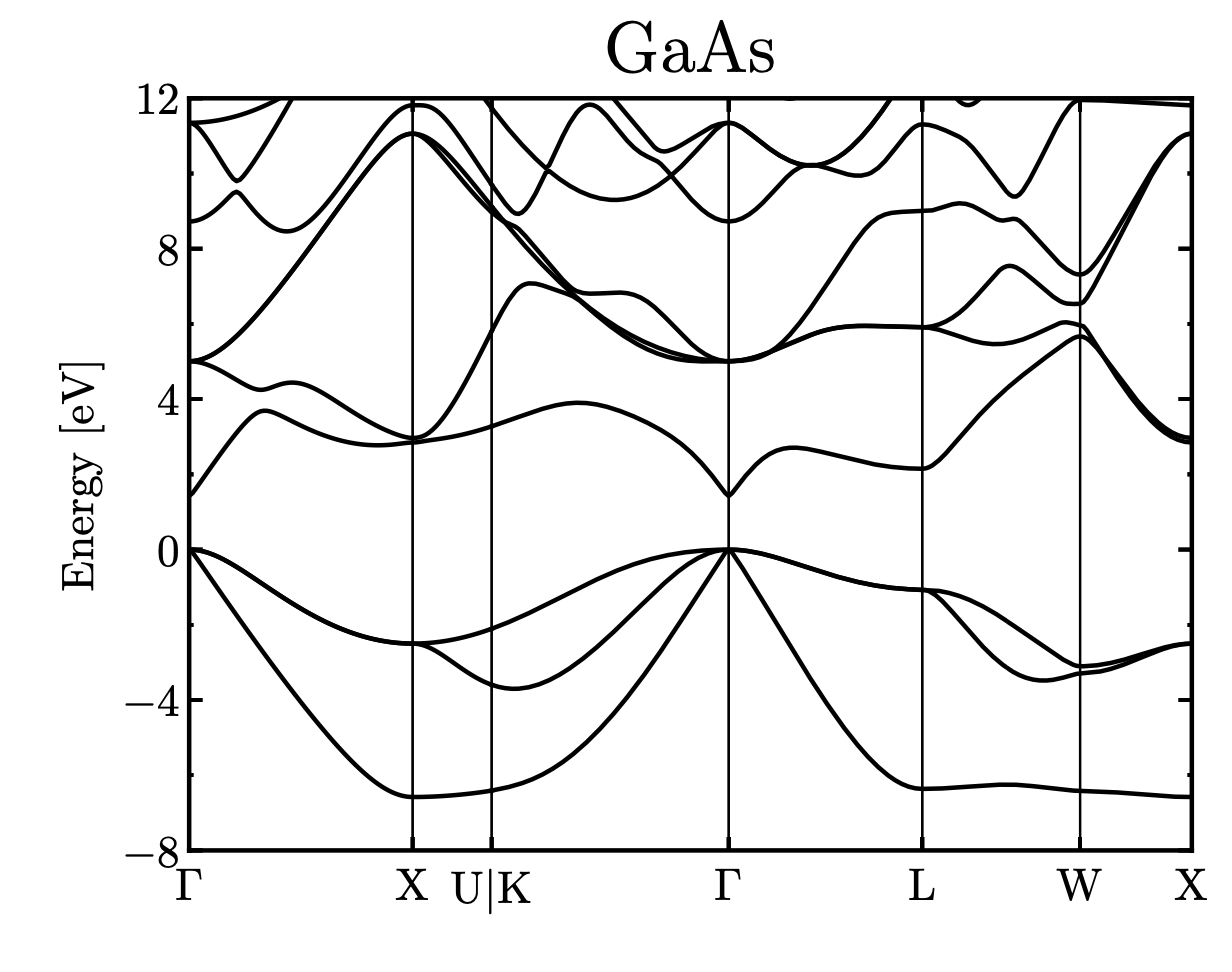}~~\includegraphics[width=0.4\textwidth]{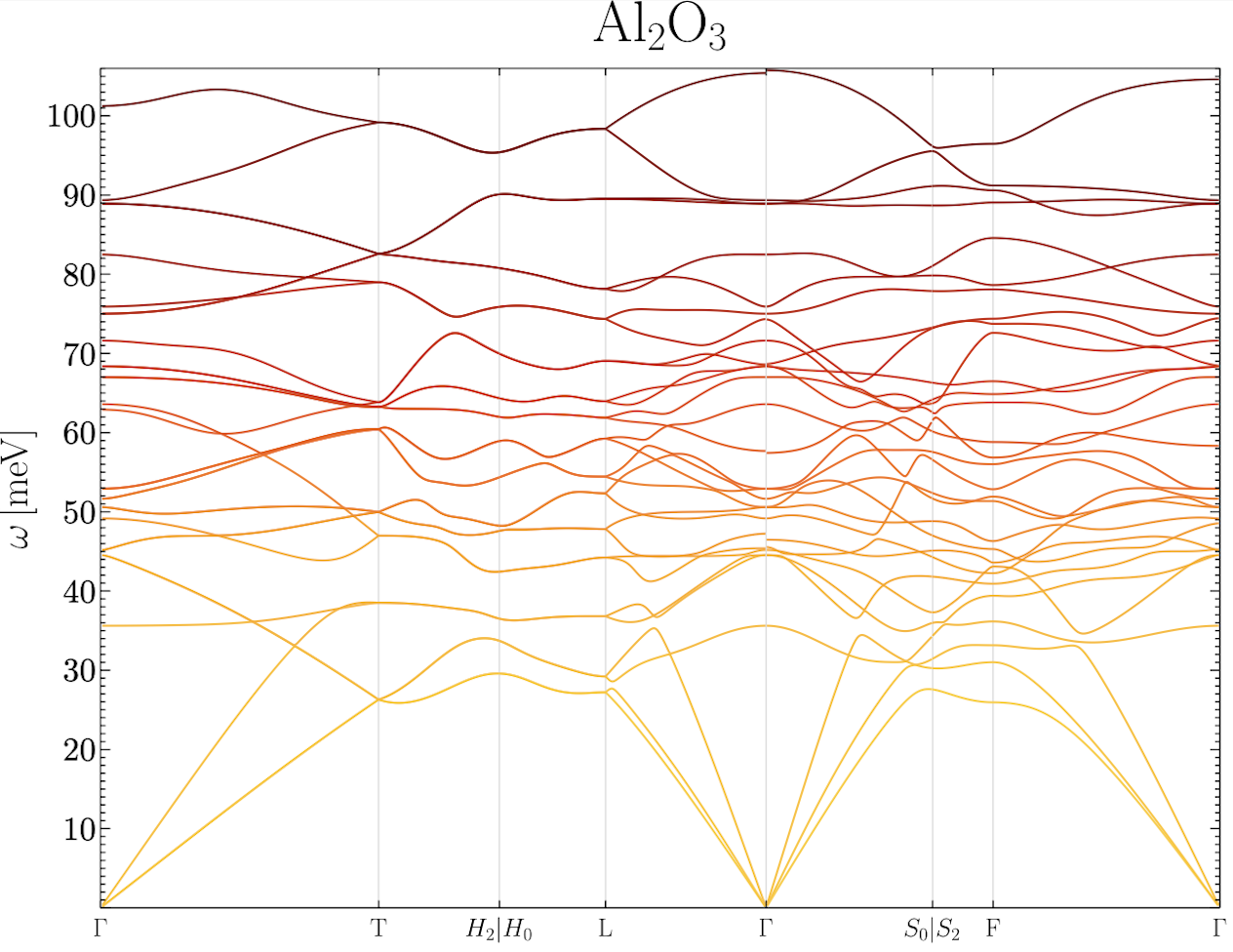}
\caption{Band structure of two materials. ({\em a}) Electronic excitations in a semiconductor, GaAs.  The Fermi energy defines the zero point on the $y$-axis, with the valence and conduction bands below and above the Fermi energy.  On the $x$-axis, a slice is taken through the Brillouin zone to show a typical band structure.  The $\Gamma$ point is defined by where the bands come closest to crossing the Fermi surface.  ({\em b}) Phonon excitations in a polar material, sapphire.  Here the excitation energies ($\omega$) are above the zero-energy state with no phonons.  The $\Gamma$ point is again defined by the point in momentum space where the gapless acoustic modes have zero energy, such that their dispersion is given by Equation~\ref{eq:gaplessphonons}.  The gapped phonons are called optical, even though not all of these modes are optically active.  Figure adapted from Reference~\cite{Griffin:2019mvc}.}
\label{fig:banddiagrams}
\end{center}
\end{figure}

When the energy deposition drops below $\omega \sim 100 \mbox{ meV}\sim q^2/2 m_N$, with $q \sim 10-100 \mbox{ keV}$, the nucleus can no longer be treated as free and the relevant degrees of freedom are no longer single ions.  Phonons are collective oscillations of atoms in fluids (such as superfluid helium) or crystals (including metals like superconductors).  For a crystal with a lattice structure having $n$ ions in a unit cell, there are $3 n$ such modes.  Three of those modes are gapless acoustic phonons; theoretically, these modes are Goldstone bosons of broken translation symmetry, and physically they correspond to the ions oscillating together in-phase in each of the three spatial directions. The dispersion of these modes is given by
 \beq
 \omega = c_s q,
 \label{eq:gaplessphonons}
 \eeq 
 where $c_s$ is the speed of sound in the medium.
Any remaining modes are gapped, meaning that at zero momentum transfer they have a non-zero excitation energy.  All of these modes physically correspond to out-of-phase oscillations of the ions, which can set up an oscillating dipole in the unit cell.  The gapped phonons are called optical phonons because at least some of these modes are optically active.  A typical band structure is shown in Figure~\ref{fig:banddiagrams}.

 Similar to the cases discussed above, the spin-independent scattering rate can be expressed in terms of a Dynamic Structure Factor, 
 \beq
 S({\bf q},\omega) =\sum_\nu \frac{|F_{\nu}({\bf q})|^2}{2 \omega_{\nu,{\bf q}}} \delta(\omega_{\nu,{\bf q}}-\omega),
 \eeq
 where $\omega_{\nu,{\bf q}}$ and $\epsilon_{\nu,{\bf q},j}$ are the eigenvalues and eigenvectors of phonon branch $\nu$ (with the polarization vector indicating the direction in which ion $j$ is oscillating, normalized such that $\sum_j  |{\bf \epsilon}_{\nu,{\bf q} j}|^2 = 1$) of the coupled ionic oscillators, and $F_{\nu}({\bf q})$ is a phonon form factor
 \beq
 |F_{\nu}({\bf q})|^2 = \left| \sum_j \frac{e^{-W_j({\bf q})}}{\sqrt{m_j}}{\bf q} \cdot {\bf \epsilon}_{\nu,{\bf q} j} e^{-i {\bf q} \cdot {\bf x}_j^0} \right|^2.
 \label{eq:phononform}
 \eeq
Here the sum runs over the ions  at equilibrium position ${\bf x}_j^0$ in the unit cell, and $W_j$ is the so-called Debye--Waller factor that acts as a form factor shutting off the phonon response when the momentum transfer becomes larger than the inverse unit cell size $q \gtrsim a^{-1}$  (for a derivation and discussion, see References~\cite{Knapen:2017ekk,Griffin:2018bjn,Trickle:2019nya}).  The eigenvalues and eigenvectors are typically obtained from a program like {\tt PhonoPy} \cite{phonopy-phono3py-JPCM,phonopy-phono3py-JPSJ} that computes the lattice force matrix on all $n$ ions in the unit cell and diagonalizes it. 
 
  Operationally, the phonon form factor behaves similarly to the Helm form factor: It becomes highly suppressed when the effective description of collectively oscillating ions (or, in the case of the Helm, collective nucleons in a nucleus) breaks down due to resolving the internal structure of the unit cell.  In particular, one can see explicitly that when the momentum transfer becomes large compared with the typical momentum in a mode, $q^2 \gg 2 \omega_{\nu, q} m_j$, the Debye--Waller factor becomes large:
 \beq
 W_j({\bf q}) = \frac{1}{4 m_j} \sum_\nu \int_{\rm 1BZ} \frac{d^3 k}{(2 \pi)^3} \frac{|{\bf q} \cdot {\bf \epsilon}_{\nu,{\bf k} j}|^2}{\omega_{\nu,{\bf k}}},
 \eeq
and hence the Dynamic Structure Factor via the phonon form factor in Equation~\ref{eq:phononform} becomes small.  (Here the momentum has been integrated over the first Brillouin zone, denoted by $1$BZ; for more detail, see Reference~\cite{Trickle:2019nya}.)
 At a general momentum transfer, the phase factors, and in particular cancellations between the phase factors, can become important for accurately describing the DM interaction rate with phonons.  However, at low momentum transfer, where the Debye--Waller factor is small, the form factor takes a simple form.  For example, in a simple crystal like GaAs in which there are only two ions in the unit cell, we have
 \beq
  |F_{\nu}({\bf q})|^2 \approx \frac{q^2}{2 m_n} \left| \sqrt{A_{\rm Ga}} e^{i {\bf x}_{\rm Ga} \cdot {\bf q}}\pm \sqrt{A_{\rm As}} e^{i {\bf x}_{\rm As} \cdot {\bf q}}\right|^2,
 \eeq
 where ${\bf x}_{\rm Ga,As}$ and $A_{\rm Ga,As}$ denote the positions and mass numbers of the gallium and arsenide ions in the unit cell, respectively~\cite{Knapen:2017ekk}.
 In general, however, one uses codes employing DFT to compute the phonon eigenvectors and eigenfrequencies~\cite{Griffin:2018bjn,Trickle:2019nya}.  This general program was outlined first in Reference~\cite{Trickle:2019nya}, and a code implementing this program in a variety of materials is publicly available as the \texttt{PhonoDark} code~\cite{Trickle:2020oki,Mitridate:2023izi}. This code not only implements spin-independent interactions but also follows a general framework the calculate the DM single-phonon excitation rate via any Lorentz-invariant effective interaction.  We describe this framework in the next section.
 
Experimental efforts are now underway to detect single collective excitations:  
 \begin{itemize}
\item The TESSERACT collaboration, consisting of the helium experiment HeRALD and the polar material experiment SPICE, is actively working to reach single optical phonon sensitivity with the transition edge sensors employed in the detector.  
\item To date, the single magnon proposal of References~\cite{Trickle:2019ovy} and \cite{Mitridate:2020kly} has not been experimentally implemented, though a related concept (through many magnons) forms the basis of the QUAX experiment~\cite{QUAX:2020adt}.
\end{itemize}
A summary of the collective modes, possible targets, gaps of each collective mode, and couplings (nucleon or electron) is given in Table~\ref{tab:PhononScatt}.  We again refer the reader to Reference~\cite{Essig:2022dfa} for a currently complete discussion of ongoing experimental efforts. 

An additional comment is in order: At typical energies above the higher optical mode (typically $\omega \sim 100 \mbox{ meV}$) but below where the nucleus becomes definitely free ($\omega \sim 1 \mbox{ eV}$), multiphonon emission becomes important.  The calculation of multiphonon processes can rapidly become numerically intensive due to the large multiphonon phase space, and the presence of both harmonic and anharmonic multiphonon modes.  Two-phonon production as a means to detect light DM was proposed in Reference~\cite{Schutz:2016tid} and \cite{Knapen:2016cue} and was calculated in an EFT in References~\cite{Caputo:2019ywq} and \cite{Acanfora:2019con}.  The harmonic contributions can be computed using analytic methods, the results can be applied to interpolate between the single phonon regime valid at low energies and the nuclear recoil regime~\cite{Campbell-Deem:2022fqm}, and the effect of anharmonicities can be estimated~\cite{Lin:2023slv}.  Note that even single phonon production will, initially, lead to the cascade of multiphonons~\cite{Baym:2020uos} as the initial phonon decays through the anharmonic coupling, in a process rather analogous to showering.
 
\begin{table}
\begin{tabular}{c|c|c|c|c}
\hline
Collective mode  & Target & Proposed materials & Gap? & Coupling \\ \hline \hline
Acoustic phonon & All materials & He, Si, Ge, GaAs, Al$_2$O$_3$, diamond & No & $p/n/e^-$ \\ \hline
Optical phonon & Polar, semiconductor & GaAs, Al$_2$O$_3$ & $\sim 10$--$100$~meV & $p/n/e^-$ \\
Magnon & (Anti-)ferromagnet & Yttrium iron garnet & $\sim0$--$10$~meV & $e^-$ \\
\hline \hline
\end{tabular}
\caption{Summary of the target materials that have been proposed for DM detection through collective excitations.  The superconductor DM detection process is elastic for energy depositions well above the Cooper pair binding energy.}\label{tab:PhononScatt}
\end{table}

\subsection{In-Medium Effects}

 Small-gap electronic materials have a large in-medium response that affects the reach to DM scattering and absorption through screening effects.  For isotropic, non-magnetic materials interacting through a dark photon, the effect of the in-medium response can be parameterized in terms of a reduced effective kinetic mixing parameter, $\epsilon_{\rm eff}$ \cite{Hochberg:2015fth,Knapen:2017xzo}:
\beq
\epsilon_{\rm eff} = \epsilon \frac{q^2}{q^2 - \Pi_{T,L}},
\label{eq:epseff}
\eeq  
where $\Pi_{T,L}$ is the in-medium polarization tensor of an isotropic, non-magnetic material related to the complex index of refraction $\tilde n$ by $q^2(1-\tilde n^2) = \Pi_L$ and $\omega^2(1-\tilde n^2) = \Pi_T$. The case of an anisotropic material is significantly more involved~\cite{Coskuner:2019odd}; the in-medium effects must be written in terms of a dielectric tensor, $\epsilon_{ij}$, whose eigenvalues are given by
\beq
\pi_i = \omega^2(1-{\bf \epsilon}_{ii}),
\eeq 
with the effective mixing parameter
\beq
\epsilon_{\mathrm{eff},i}^2 = \frac{\epsilon^2 m_{A'}^4}{\left[m_{A'}^2 - \mbox{Re} \pi_i(q)\right]^2 + \left[\mbox{Im} \pi_i(q) \right]^2}.
\eeq
We discuss below how these effective couplings enter into the absorption rate.

\section{Generalized Dark Matter Interactions}


We have summarized the relevant ingredients for DM to induce a response in a target material.  We now generalize the framework to an EFT of DM scattering in Section~\ref{sec:EFT}, followed by general considerations regarding DM absorption in target materials.

\subsection{Effective Field Theory of Dark Matter Scattering with Collective Excitations}
\label{sec:EFT}

 As suggested in Equation~\ref{eq:dynamic}, one needs to be able to compute the transition matrix element $\langle f | {\cal F}_T({\bf q})|i \rangle$ for any interaction type in order to compute the DM scattering rate for any interaction type.  This, in turn, implies that we must be able to compute the potential---the generalization of Equation~\ref{eq:potential}---that the DM induces in the target material.   DM is a nonrelativistic state, and therefore one needs to follow the rules of non-relativistic EFT (NREFT).  The discussion here largely follows Reference~\cite{Trickle:2020oki}. 

For scattering, the EFT calculation follows a simple plan:
\begin{enumerate}
\item Match relativistic operators onto non-relativistic operators.  There is a long history in the nuclear physics literature of identifying the relevant nonrelativistic operators.  They are
\beq
{\bf q}\equiv {\bf k}'- {\bf k},~~~~{\bf K} \equiv {\bf k}'+ {\bf k},~~~~{\bf S}_\psi,~~~{\bf v}_\chi \equiv \frac{ {\bf P}}{2 m_\chi},~~\text{and}~~{\bf v}_\psi \equiv \frac{ {\bf K}}{2 m_\psi}.
\label{eq:NRops}
\eeq
Here, $m_\psi$ is the target fermion mass and ${\bf S}_\psi$ is the spin.  Note that in the nuclear recoil case, Galilean invariance is preserved, and the NREFT depends only on ${\bf v}^\perp \equiv {\bf v}_\chi - {\bf v}_\psi$.  In the present case, in-medium effects may be important, and Galilean invariance is broken such that we need to keep both.
\item Equation~\ref{eq:NRops} identifies charge, spin, and velocity operators as relevant to  determining the potential created by the DM--target interaction.  In particular, the transition matrix element induced by a potential, generalized from Equation~\ref{eq:potential} to include velocity and momentum dependence, is
\beq
\langle \nu,{\bf k} | \tilde V({\bf q},{\bf v})| 0 \rangle = \sum_{\ell,j} \langle \nu,{\bf k} | e^{- i {\bf q} \cdot {\bf x}_{\ell j}} \tilde V_{\ell j}({\bf q},{\bf v})| 0 \rangle , 
\eeq
where $\ell$ labels the unit cell, $j$ denotes the ion within the unit cell, and the subscripts on the potential denote the contribution from each lattice site:
\beq
V({\bf x},{\bf v}) = \sum_{\ell j} V_{\ell j}({\bf x} - {\bf x}_{\ell j},{\bf v}),
\eeq
with 
\beq
\tilde V({\bf q},{\bf v}) = \int d^3 x e^{- i {\bf q} \cdot {\bf x}} V({\bf x}, {\bf v}). 
\label{eq:qspacepot}
\eeq
The transition is between the ground state of the material $| 0 \rangle$ and a state with a single collective excitation, labeled $| \nu, {\bf k} \rangle$.   Because in what follows we exclusively use the potential in momentum space, we hereafter drop the tilde on $\tilde V({\bf q},{\bf v})$ for readability.
\item Finally, we quantize the lattice potential to compute the matrix element.  There are two quanta---phonons and magnons---that are excitations of the lattice potential that we consider.  
\begin{itemize}
\item Phonons are quanta of lattice displacement ${\bf u}_{\ell j}$ from the equilibrium ion position at site $j$ in the $\ell$th unit cell ${\bf x}_{\ell j}^0$:
\beq
{\bf u}_{\ell j} = {\bf x}_{\ell j} - {\bf x}_{\ell j}^0 = \sum_{\nu = 1}^{3 n} \sum_{\bf k} \frac{1}{\sqrt{2 N m_j \omega_{\nu,k}}} \left(\hat a_{\nu,k} {\bf \epsilon}_{\nu,{\bf k} j} e^{i {\bf k} \cdot {\bf x}_{\ell j}^0} +  \hat a_{\nu,k}^\dagger  {\bf \epsilon}^*_{\nu,{\bf k} j} e^{-i {\bf k} \cdot {\bf x}_{\ell j}^0}\right),
\eeq
where $N$ is the number of cells in the lattice, $m_j$ is the mass of the ion at the $j$th site, and we have added subscripts on the energy deposition $\omega_{\nu,{\bf k}}$ (where the momentum ${\bf k}$ is in the first Brillouin zone) to emphasize that the energy deposition must correspond to the energy of one of the eigenfrequencies of the collective excitations.
 Since the matrix element we seek to compute involves the potential from Equation~\ref{eq:qspacepot} with a factor of $e^{i {\bf q} \cdot {\bf x}_{\ell j}}$, the matrix element must be evaluated via Campbell--Baker--Hausdorff to give\footnote{the derivation can be found in Reference~\cite{Trickle:2019nya}. }
\beq
\langle \nu,{\bf k}| e^{-i {\bf q} \cdot {\bf x}_{\ell j}}  V_{\ell j}({\bf q},{\bf v}) | 0 \rangle = \frac{1}{\sqrt{V}} \sum_{\nu,{\bf k}j} \left[\sum_\ell  V_{\ell j}({\bf q},{\bf v}) e^{i ({\bf q} - {\bf k}) \cdot {\bf x}_{\ell j}^0 }  \right] \frac{e^{-W_j}({\bf q}) ({\bf q} \cdot {\bf \epsilon}^*_{\nu,{\bf k}j})}{\sqrt{2 m_j \omega_{\nu,{\bf k}}}}. \nonumber
\eeq
The task is then to evaluate the lattice potential, $V_{\ell j}({\bf q},{\bf v})$.  This depends on the nature of the interaction.  Using an example from Reference~\cite{Trickle:2020oki}, involving four different types of responses, a lattice potential may take the following form: 
\begin{eqnarray}
V_{\ell j}({\bf q},{\bf v})  \supset & \sum_\alpha &  \left[ c_1 \langle e^{i {\bf q} \cdot {\bf x}_\alpha} \rangle_{\ell j} + c_4 {\bf S}_\chi \cdot \langle e^{i {\bf q} \cdot {\bf x}_\alpha} {\bf S}_\alpha \rangle_{\ell j} \right. \\ \nonumber
 & &  \left. + c_{8b} {\bf S}_\chi \cdot \langle e^{i {\bf q} \cdot {\bf x}_\alpha} {\bf v}_\alpha \rangle_{\ell j} + c_{3b}\frac{i {\bf q}}{m_\psi} \cdot \langle e^{i {\bf q} \cdot {\bf x}_\alpha} {\bf v}_\alpha \times {\bf S}_{\alpha}\rangle_{\ell j}
 \right].
\end{eqnarray}
A spin-independent interaction requires one to evaluate the matrix element
\beq
\sum_\alpha \langle e^{i {\bf q} \cdot {\bf  x}_\alpha} \rangle_{\ell j} \simeq \langle N_\psi \rangle_{\ell j}, 
\eeq
where $\alpha$ runs over fermions of type $\psi = p,~n,~e$.  Likewise, a spin-dependent interaction requires one to evaluate
\beq
\sum_\alpha \langle e^{i {\bf q} \cdot {\bf  x}_\alpha} {\bf S}_{\psi,\alpha} \rangle_{\ell j} \simeq \langle {\bf S}_\psi \rangle_{\ell j}.
\eeq
Coupling to electric and magnetic dipoles, as well as the anapole operator, involves the velocity ${\bf v}_{\psi,\alpha} = -\frac{i}{2 m_\psi} \overleftrightarrow{\nabla} $, and one must evaluate the expectation value, 
\beq
\sum_\alpha \langle e^{i {\bf q} \cdot {\bf x}_\alpha} {\bf v}_{\psi,\alpha} \rangle_{\ell j} \simeq i \langle ({\bf q} \cdot {\bf x}_\alpha) {\bf v}_{\psi,\alpha} \rangle_{\ell j},
\eeq 
which becomes
\beq
\langle x^i_\alpha v_{\psi,\alpha}^k \rangle_{\ell j} = -\frac{i}{2 m_\psi} \langle x^i  {\bf \nabla}_\alpha^k - x^k  {\bf \nabla}_\alpha^i \rangle_{\ell j} = \frac{1}{2 m_\psi} \epsilon_{i k k'} \langle L_\alpha^{k'} \rangle_{\ell j},
\eeq
where $L$ is an angular momentum operator.
Overall, one finds in the long-wavelength limit that there are four types of responses:
\beq
N,~~S,~~L,~\text{and}~L\otimes S,
\eeq
where we have not written out the decomposition of the last operator because it does not commonly appear in Lorentz-invariant ultraviolet-completions.

\item Magnons are quanta of spin precession.  Here, the relevant matrix element is
\beq
\langle \nu,{\bf k}| V({\bf q},{\bf v})|0 \rangle = \sum_{\ell,j} e^{i {\bf q} \cdot {\bf x}_{\ell j}} {\bf f}_j({\bf q}, {\bf v}) \cdot \langle \nu,{\bf k} | {\bf S}_{\ell j} | 0 \rangle,
\eeq
where ${\bf f}_j$ is dependent on the interaction type, and ${\bf S}_{\ell j}$ denotes the ion effective spins, which can come both from electronic spin and orbital angular momentum (for details, see References~\cite{Trickle:2020oki} and \cite{Trickle:2019ovy}).  This shows that one needs a net spin on each unit cell to excite a response.
\end{itemize}
\end{enumerate}

The rate is then computed from Fermi's Golden Rule (Equation~\ref{eq:FGR}), which in the case at hand becomes
\beq
\Gamma({\bf v}) = \int \frac{d^3 q}{(2 \pi)^3} \sum_{\nu,{\bf k}} \left|\sum_{\ell,j} \langle \nu,{\bf k}| e^{-i {\bf q} \cdot {\bf x}_{\ell j}} V_{\ell j}({\bf q},{\bf v}) | 0 \rangle \right|^2 2 \pi \delta(E_f - E_i - \omega).
\eeq


\subsection{Target Response to Dark Matter Absorption}

In this subsection, we discuss DM absorption in materials, and focus on the case of vector DM and pseudoscalar (axion) DM.  For the case of absorption, the energy absorbed is simply the mass, $\omega = m_\chi$, which is much greater than the momentum, $q = m_\chi v$, $\omega \gg q$.  This implies one of two possibilities to kinematically allow for DM abosrption:
\begin{enumerate}
\item an inelastic transition in the target material, implying the presence of a gapped mode having $\omega_q \neq 0$ as $q \rightarrow 0$; or
\item an absorption process with two excitations in the final state, where the momenta of the two excitations cancel (to high precision) while the sum of their energies equals $m_\chi$.
\end{enumerate}

It has long been appreciated that new particles (produced in the Sun) can be absorbed on  target materials via inelastic transitions~\cite{Dimopoulos:1986mi,Gelmini:1987nx}, such as valence electrons in semiconductors making a valence-to-conduction-band transition.  More recently, these ideas were applied to vector axion DM absorption on electrons in xenon~\cite{Pospelov:2008jk,An:2014twa}.  The absorption rate can be related to the complex conductivity via the optical theorem
 \beq
\Gamma_\gamma = - \frac{\mbox{Im}\Pi(\omega)}{\omega},
 \label{eq:pi}
 \eeq
 where $\Pi$ is related to the complex conductivity $\hat \sigma(\omega)$:
 \beq
 \Pi(\omega) \approx - i \hat \sigma \omega,
 \eeq
 with $\Pi$ having transverse and longitudinal polarizations as in Equation~\ref{eq:epseff}.
Since $|{\bf q}| \ll \omega$, $\Pi_L = \Pi_T \simeq \Pi$.  The dark photon absorption rate, per unit of target mass, is given by 
\beq
\Gamma_{A'} = \frac{1}{\rho_T}\frac{\rho_\chi}{m_\chi} \epsilon_{\rm eff}^2 \Gamma_\gamma,
\label{eq:photontodarkphoton}
\eeq
with $\epsilon_{\rm eff}$ (for an isotropic non-magnetic medium) given by Equation~\ref{eq:epseff}.
For an anisotropic material, the absorption rate, per unit of target mass, is as follows~\cite{Coskuner:2019odd}:
\beq
R = -\frac{1}{3} \frac{\rho_\chi}{\rho_T} \sum_{i = 1}^3 \epsilon_{\mathrm{eff},i}^2\frac{\mbox{Im}\pi_i(q)}{m_{A'}^2},
\eeq
where $\pi_i$ denotes the eigenvalues of the polarization tensor.

The axion absorption rate on electrons can be extracted from Equation~\ref{eq:pi} by using the relation with the photon absorption rate~\cite{Hochberg:2016ajh}:
\beq
|{\cal M}|^2 \approx 3 (g_{aee}/2 m_e)^2 (m_a/e)^2|{\cal M}_\gamma|^2.
\label{eq:photontoaxion}
\eeq
Axions can also be absorbed on gapped optical phonons~\cite{Mitridate:2020kly}.  These modes, similar to electrons in a semiconductor, have a gap at zero momentum transfer (see Figure~\ref{fig:banddiagrams}).  Unlike the case of absorption on electrons, where one can make use of the direct axion electron coupling, the interaction goes via the mixing of the phonon with the photon (known as the phonon polariton), in the presence of an external $B$ field.
  
In the second process enumerated above, two modes recoil against each other.  In this case, the momentum can be conserved by a cancellation between the two outgoing modes; this cancellation is necessary for acoustic phonons because, for a fixed energy deposition, they have a large amount of momentum compared with the DM momentum due to the small speed of sound compared with the DM velocity, $c_s \ll v_\chi$: 
\beq
q_{\rm ph} = \frac{\omega}{c_s}  \gg \frac{\omega}{v_\chi}.
\eeq
For example, bosonic DM can be absorbed on free electrons in superconductors by emitting a phonon~\cite{Hochberg:2016sqx}, on two gapless phonons in superfluid helium~\cite{Knapen:2016cue}, or on a photon and phonon~\cite{Murgui:2022zvy}.  All become kinematically possible because of the back-to-back recoil of the two final-state excitations.  Note that in the cases where only one excitation is produced ({\em e.g.} electron or optical phonon excitation in semiconductors), momentum is conserved by recoil against the lattice.

\begin{figure}
\begin{center}
\includegraphics[width=0.8\textwidth]{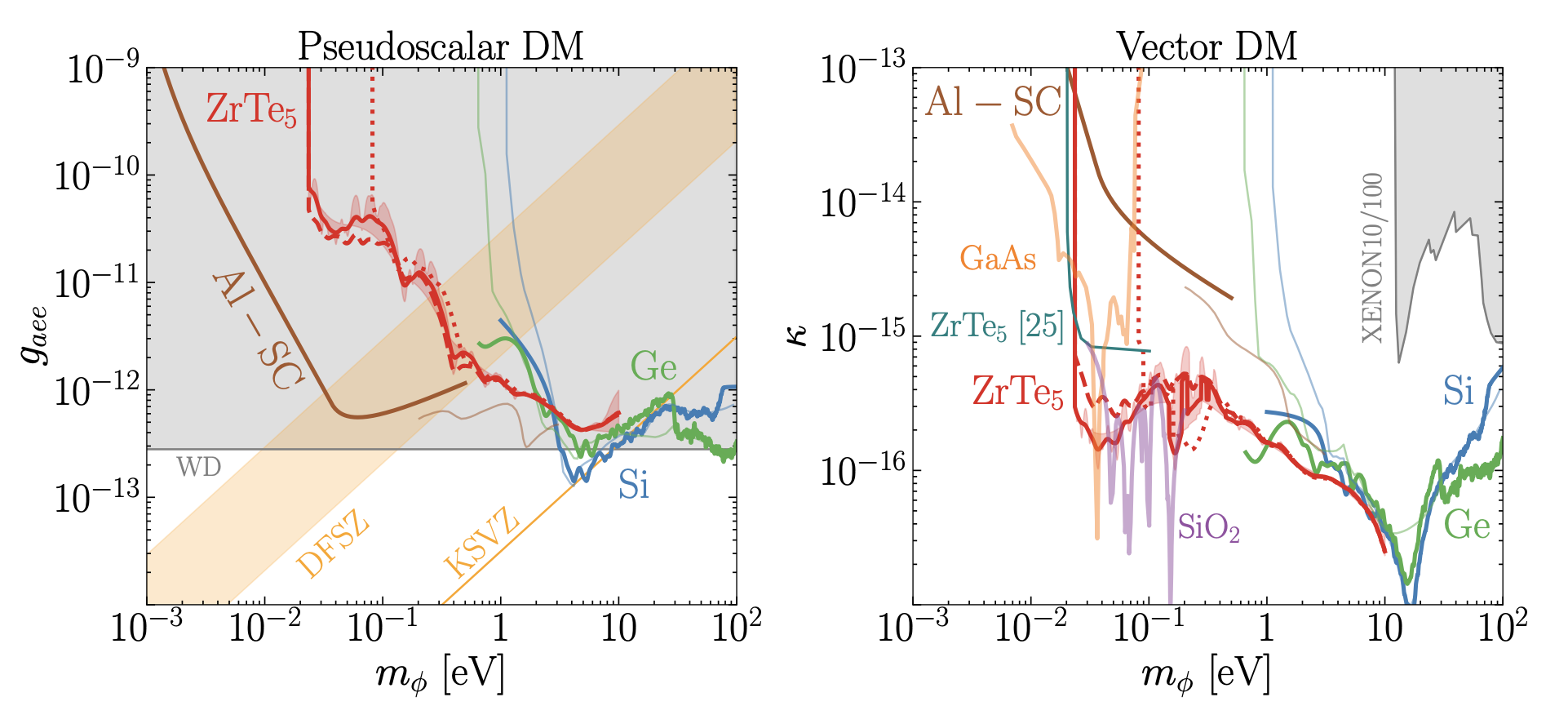}  \includegraphics[width=0.5\textwidth]{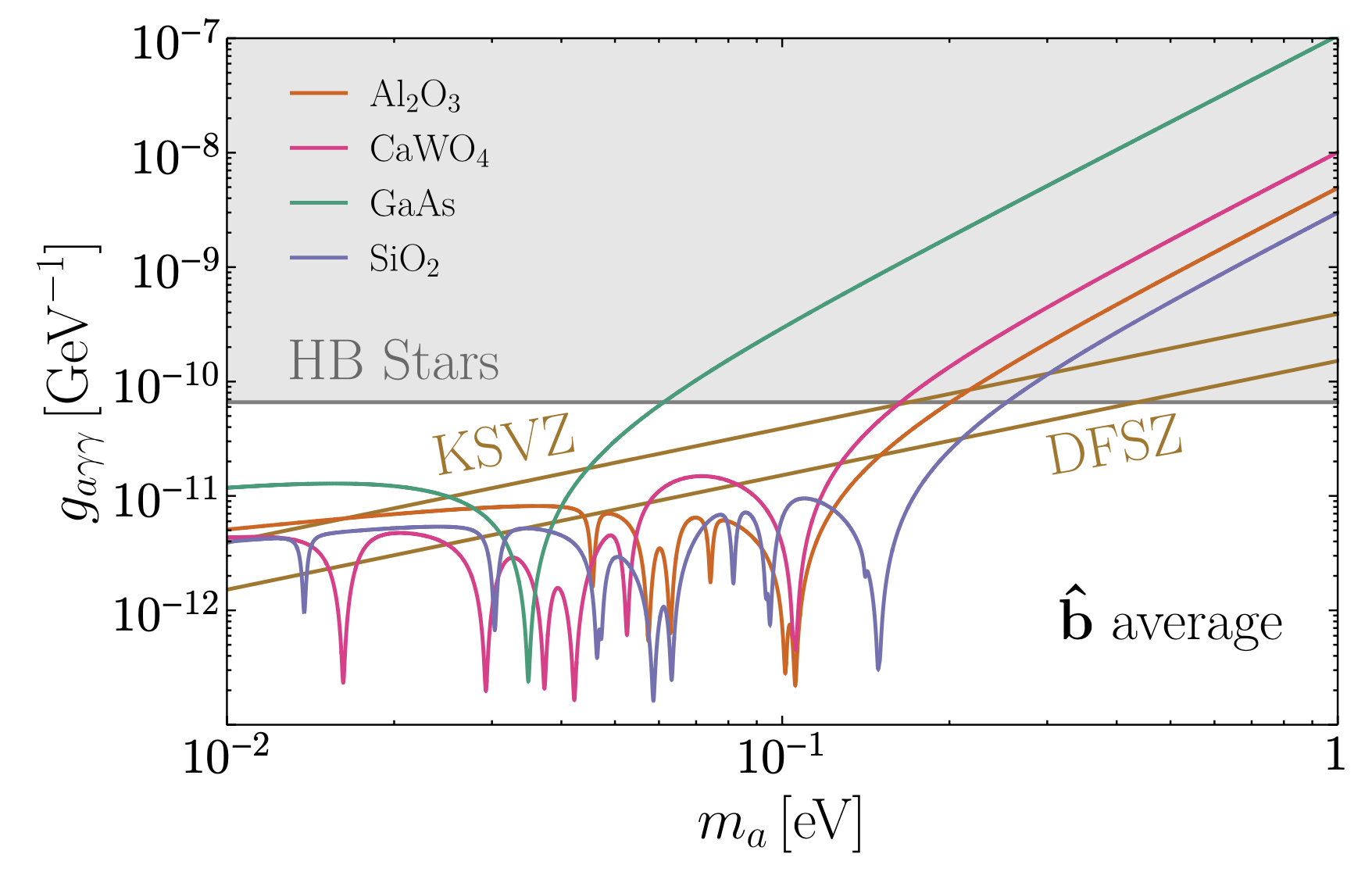}
\caption{{\em Upper}: Reach to {\em (a)} axion DM and {\em (b)} vector DM via their {\em (a)} electron coupling or {\em (b)} kinetic mixing coupling to the target material, for $1$-kg-year exposure.  {\em (c)} Reach to axion DM by absorption on phonon polaritons in a $1$-T external magnetic field. Abbreviations: Al SC, superconducting aluminum; DM, dark matter; HB, horizontal branch; WD, white dwarf. Panels {\em a} and {\em b} adapted from Reference~\cite{Chen:2022pyd}. Panel {\em c} adapted from Reference~\cite{Mitridate:2020kly}.}
\label{fig:absorptionreach}
\end{center}
\end{figure}

 A summary plot comparing the reach of axion and dark photon absorption on electrons in semiconductors and superconductors, and on phonon polaritons in polar crystals, is shown in Figure~\ref{fig:absorptionreach}.
 
Here, we have chosen to restrict ourselves to particular models whose absorption rate can be simply related to photon interaction rates.  One can also pursue an EFT framework for absorption on both electrons~\cite{Chen:2022pyd,Krnjaic:2023nxe} and collective modes (phonons~\cite{Mitridate:2023izi} and magnons~\cite{Trickle:2019ovy}).  We direct the reader to the references cited here for further details on more general types of interactions, including DM absorption via electric and magnetic dipoles.

\section{Conclusion}

We have reviewed the development of theories of particle DM of a very low mass---below the traditional WIMP window of $\sim 10$~GeV but above masses of $\sim 1$ eV, where DM becomes wavelike ({\em e.g.} axions).  Such models are motivated by hidden-sector theories and have rich cosmological and astrophysical dynamics, from self-interactions to impacts on stellar evolution and observational consequences in collider experiments.  We have reviewed the astrophysical, cosmological and collider constraints most directly relevant for the model space in terrestrial direct detection experiments.  

While 15 years ago, direct detection experiments could not reach the sub-GeV DM mass window, the proposal of HVDM/HSDM led to an explosion of ideas for direct detection experiments.  A subsequent push in the last 5-10 years to realize these experiments with research and development gave rise to funded experiments that are actively reaching new theory space.  At the present moment, these experimental efforts have not yet covered the best-motivated candidates, such as asymmetric DM, thermal freeze-out DM, and DM produced through freeze-in.  As these new experiments come to fruition and push to lower cross sections with better handles on backgrounds and systematic uncertainties, we look forward to the possible uncovering of the Universe's dark side.

\section*{Disclosure Statement}

The author is not aware of any affiliations, memberships, funding, or financial holdings that might be perceived as affecting the objectivity of this review.

\acknowledgments

I thank Yufeng Du, Osmond Wen, Zhengkang Zhang, and especially Clara Murgui and Tanner Trickle for a careful reading of the manuscript.  This work was supported by the US Department of Energy, Office of Science, Quantum Information Science Enabled Discovery (QuantISED) for High Energy Physics (KA2401032), by the Office of High Energy Physics (award DE-SC0011632), by a Simons Investigator Award, and by the Walter Burke Institute for Theoretical Physics.

\bibliography{reviewbib.bib}

\end{document}